\documentclass[12pt,preprint]{aastex}
\usepackage{psfig,lscape}
\def\lap{\lower.5ex\hbox{$\; \buildrel < \over \sim \;$}}
\def\gap{\lower.5ex\hbox{$\; \buildrel > \over \sim \;$}}

\def\ergcm2s{${\rm erg\ cm^{-2}\ s^{-1}}$}

\def\ergscm2s{${\rm erg\ cm^{-2}\  s^{-1}}$}

\def\cm-2{${\rm cm^{-2}}$}
\def\ergs{${\rm erg\ s^{-1}}$}

\begin{document}

\title{A Catalog of Transient X-ray Sources in M31}

\author{Benjamin~F.~Williams\altaffilmark{1,2}, S.~Naik\altaffilmark{3,4}, Michael~R.~Garcia\altaffilmark{1}, Paul J. Callanan\altaffilmark{3}}
\altaffiltext{1}{Harvard-Smithsonian Center for Astrophysics, 60
Garden Street, Cambridge, MA 02138; williams@head-cfa.harvard.edu;
garcia@head-cfa.harvard.edu} 
\altaffiltext{2}{Department of Astronomy and Astrophysics, 525 Davey Lab, Pennsylvania State University, University Park, PA 16802; bwilliams@astro.psu.edu}
\altaffiltext{3}{Department of Physics, University College Cork, Cork,
Ireland; sachi@ucc.ie; paulc@ucc.ie}
\altaffiltext{4}{Institute of Space and Astronautical Science, Japan Aerospace Exploration Agency, 3-1-1 Yoshinodai, Sagamihara, Kanagawa 229-8510, Japan; naik@astro.isas.jaxa.jp}


\begin{abstract}

From October 1999 to August 2002, 45 transient X-ray sources were
detected in M31 by {\it Chandra} and {\it XMM-Newton}.  We have
performed spectral analysis of all {\it XMM-Newton} and {\it Chandra}
ACIS detections of these sources, as well as flux measurements of {\it
Chandra} HRC detections.  The result is absorption-corrected X-ray
lightcurves for these sources covering this 2.8 year period, along
with spectral parameters for several epochs of the outbursts of most
of the transient sources.  We supply a catalog of the locations,
outburst dates, peak observed luminosities, decay time estimates, and
spectral properties of the transient sources, and we discuss
similarities with Galactic X-ray novae.  Duty cycle estimates are
possible for 8 of the transients and range from 40\% to 2\%; upper
limits to the duty cycles are estimated for an additional 15
transients and cover a similar range.  We find 5 transients which have
rapid decay times and may be ultra-compact X-ray binaries.  Spectra of
three of the transients suggest they may be faint Galactic foreground
sources.  If even one is a foreground source, this suggests a surface
density of faint transient X-ray sources of $\gap$1 deg$^{-2}$.

\end{abstract}

\section{Introduction}

Bright X-ray transient (XRT) events provide brief glimpses into the
physical state of accreting compact objects.  Detailed studies of
Galactic soft XRTs, also known as X-ray Novae (XRNe), have put
constraints on the physics at work in these systems
(e.g. \citealp{chen1997}).  Optical follow-up of 15 Galactic XRNe has
revealed that the systems each contain an accreting object more
massive than $\sim 3$ M$_{\odot}$ \citep{mcclintock2003}, which cannot
be a stable neutron star (NS).  Their high X-ray luminosities
($>10^{38}$\ergs\/) and millisecond X-ray variability further support
the argument that these binary systems contain black holes (BHs).
Therefore, these objects are among the most secure examples of black
holes known \citep{charles1998}.

Galactic XRTs occur as both high-mass X-ray binaries (HMXBs) and
low-mass X-ray binaries (LMXBs). The secondary stars of outbursting
HMXBs are typically B/Be type, and the primaries are typically neutron
stars \citep{tanaka1996}. These HMXBs tend to exhibit pulsations, hard
spectra, and periodically repeating outbursts.

At the same time, outbursting LMXBs have often been found to contain
black hole (BH) primaries.  As black holes do not have a physical
surface, X-rays from the black hole X-ray novae (BHXNe) do not pulse.
In addition, the spectra of BHXNe are typically softer than those of
HMXBs (see the review by \citealp{mcclintock2003}).

Outside the Milky Way, nearby galaxies are observed to have XRTs which
are likely to be similar to those in our Galaxy.  However, their
greater distance makes it difficult to find optical counterparts and
hinders efforts to decipher detailed information about their X-ray
properties.

On the other hand, there are some advantages to extragalactic studies
of XRNe.  For example, in the case of M31, all of the bulge can be
observed in one Chandra or XMM observation, and each such observation
often reveals a new XRT source (e.g. \citealp{williams2004}).  These
observations provide accurate positions for followup observations as
well as spectral information for constraining the physical nature of
the sources during these accretion events.

Because M31 is the nearest neighboring spiral galaxy, transient X-ray
sources detected in M31 have been well-reported.  The earliest
transient source was discovered by \citet{trinchieri1991}.  Many
others followed, as observations of M31 by {\it Chandra} and {\it XMM-Newton}
have been combed in order to catalog events that resemble Galactic
XRNe.

\citet{white1995} reported a supersoft repeating transient
source. \citet{atel79} reported two more new transient sources.  Five
additional sources were seen by \citet{trudolyubov2001} and
\citet{osborne2001} in {\it XMM-Newton} and {\it Chandra} observations
from the year 2000.  \citet{iauc7659} reported
another. \citet{kong2002} cataloged 10 new X-ray transients in their
Chandra-ACIS study of the central region of M31.  \citet{iauc7798}
reported 3 more transient events from {\it XMM-Newton} data. Recently,
\citet{williams2004} cataloged 8 additional transient sources from a
{\it Chandra}-HRC survey.  Finally, \citet{distefano2004} have
completed a survey for supersoft X-ray sources (SSS) in four regions
of M31, finding 12 previously unpublished transient objects and noting
6 objects in the $ROSAT$ catalog that were undetected by $Chandra$ or
$XMM$.  Herein, we identify 2 previously-detected objects as transient
candidates.  All of these reports add up to 51 X-ray transients, 45 of
which were detected by {\it XMM-Newton} or {\it Chandra} over a 3 year
period.  These 45 are presented in Table~\ref{trans.tab}.

These transient sources in M31 offer the possibility of studying a
statistical sample of sources all at a well-known distance (780 kpc;
\citealp{williams2003sfh}) and covering only a small part of the sky.
In addition, while portions of our Galaxy cannot be effectively
surveyed in low-energy X-rays due to absorption, the M31 sample does
not suffer from this problem.  The M31 sample allows comparative
studies with the Galactic sample and adds to the number of transient
sources known.  Finally, M31 could harbor X-ray transient sources with
uncommon properties not yet discovered in the Galaxy.  Only through
detailed studies of the extragalactic siblings of Galactic XRNe will
be able to obtain a complete understanding of this class of X-ray
sources.

In order to fully exploit the available X-ray data for M31 transients
that were detected in 1999--2002, we have measured the fluxes and,
where possible, the spectra of every {\it XMM-Newton} and {\it
Chandra} detection of these 45 transient X-ray
sources. Section~\ref{data} describes the data reduction technique for
detecting and extracting the sources, measuring their fluxes and
fitting their spectra.  Section~\ref{results} details our results,
including lightcurves, spectral parameters, and classification of the
different types of spectra displayed by the transient sources.
Section~\ref{discussion} compares the properties of these events with
Galactic XRNe, and finally Section~\ref{conclusions} summarizes our
conclusions.

\section{Data Reduction and Analysis}\label{data}

We obtained archival {\it Chandra} HRC and ACIS data of M31 with
sampling every few months from October 1999 to August 2002.  In total,
81 HRC-I observations, 23 ACIS-I observations, and 17 ACIS-S
observations were included.  In addition, we obtained 4 epochs of {\it
XMM-Newton} data covering the central region of M31 and a single epoch
of {\it XMM-Newton} data covering most of the M31 disk.  Each of these
{\it Chandra} and {\it XMM-Newton} observations contained at least one
transient X-ray source, as well as upper-limit information for
transients that appeared either before or after a given observation.
The dates, instruments, observation identification (OBSID) numbers,
hardness ratios, and spectral parameters for all of the {\it
XMM-Newton} and {\it Chandra} ACIS observations are provided in
Table~\ref{spec.dat}.  The HRC observations are described in
\citet{williams2004}.

\subsection{Chandra Data}

\subsubsection{Source Detections}\label{detections}

The ACIS data were processed using the CIAO 3.1 data analysis package
with the calibration data set CALDB 2.28.  We removed the effects of
the ACIS-I charge transfer inefficiency using the task {\it
acis\_process\_events}.  We then created two images from each ACIS
observation of M31.  The first image was a full-resolution (0.5$''$
pixel$^{-1}$) image of an area 8.5$'$ X 8.5$'$ centered on the
aimpoint of the observation.  The second image was a 4$''$
pixel$^{-1}$ resolution image of the entire Chandra field.  Using
these two images allowed us to obtain more accurate positions and
avoid blending where the Chandra PSF is very good while also allowing
us to search the entire field for transient detections.  We searched
each of these images for sources in all of the {\it Chandra} ACIS data
using the CIAO task {\it wavdetect}, which created source lists for
all of the observations.

The {\it wavdetect} source lists that we generated and the {\it
celldetect} source lists that were generated for each data set by the
ACIS data reduction pipeline were searched for objects within 4$''$ of
the known transient sources, and any detections were cataloged to be
used as part of the transient lightcurve.  If no source was detected
at the transient position, the 4$\sigma$ detection limit at the
transient position in the image was calculated from the exposure and
background level at that location in the image.  The upper-limit was
converted to a luminosity using the same procedure as was used for the
{\it XMM-Newton} upper-limits, described below (\S~\ref{xmm}).
Hardness ratios were calculated for each detection using
background-subtracted counts (uncorrected for absorption or instrument
response) in three energy bands: S (0.3--1 keV), M (1--2 keV), and H
(2--7 keV).  The formulas used were $HR1 = (M-S)/(M+S)$ and $HR2 =
(H-S)/(H+S)$.

In addition, the source lists were searched for any bright sources
that were not detected in all observations and were not in the list of
published transients.  By this method, two new transient sources,
r1-28 and r3-8, were added to the catalog \citep{kong2002}.  These
sources have been cataloged previously in the literature, but were not
considered as transient sources.  Our analysis suggests that their
lightcurves are comparable to most other sources previously classified
as transients.

Finally, the X-ray fluxes from {\it Chandra}-HRC detections of the
transient events were taken from \citet{williams2004}. See
\citet{williams2004} for the full details of the HRC analysis and
results.  Herein, we have converted each HRC flux measurement to an
unabsorbed 0.3--7 keV luminosity by assuming the spectral parameters
of the most contemporaneous ACIS or {\it XMM-Newton} spectral fit.

\subsubsection{Spectral Fits}~\label{fits}

We extracted the spectrum of each transient detection that contained
enough counts ($\gap$50) using the CIAO routine {\it psextract}, so
that each bin in the spectrum contained $\gap$10 counts (more for
high-count detections).  By using CIAO 3.1 to extract the spectra, the
degradation of the ACIS quantum efficiency over time was taken into
account in the spectral extraction.  These spectra were then fit using
CIAO 3.1/Sherpa \citep{freeman2001}.  Because many of our spectra
contained only $\sim$5--10 data points, our goal in fitting the
spectra was to find the simplest model that provided a good fit to the
data.  We therefore started by trying to fit a power-law, then moved
on to more complex spectra as necessary.

\begin{figure}
\centerline{\psfig{figure=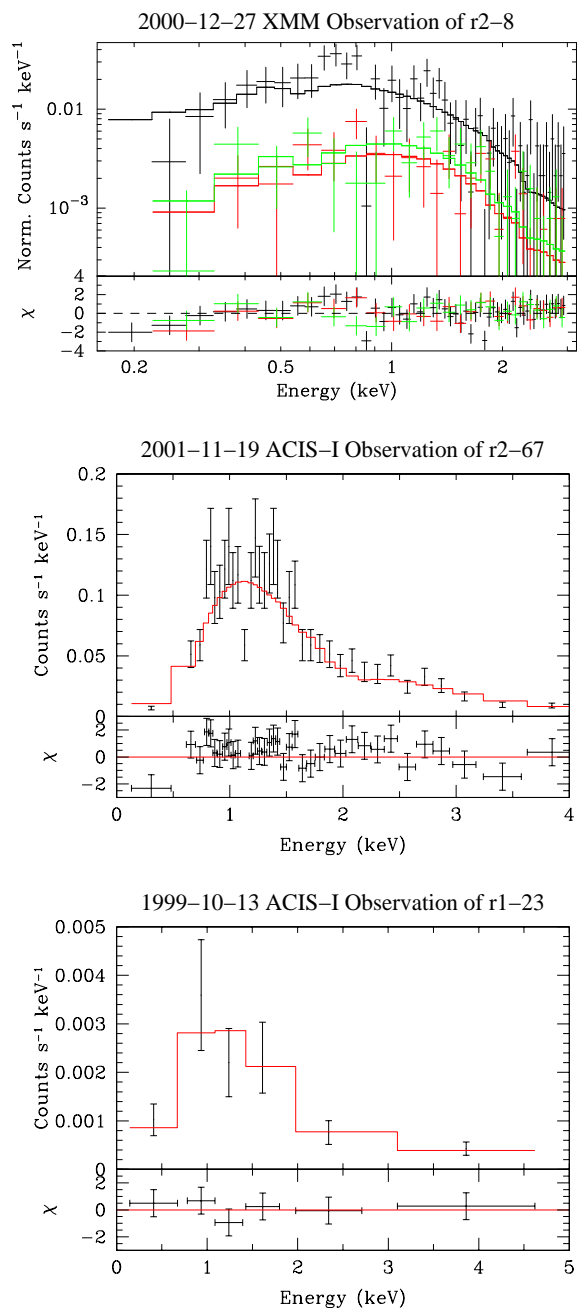,height=7in,angle=0}}
\caption{A few sample spectral fits are shown.  The top plot shows a
typical XMM fit.  The green and red spectral fits are for the MOS-1
and MOS-2 detectors, and the black fit is for the PN detector.  The
middle plot shows a typical ACIS fit for a high count rate detection,
and the bottom plot shows a typical ACIS fit for a low count rate
detection.  In all plots, histograms mark the best fit model
prediction and error bars mark the measured count rates from the
data.}
\label{spec.examples}
\end{figure}

In cases where a good fit ($\chi^2/\nu < 1.2$) could not be found
using an absorbed power-law, an absorbed blackbody thermal model and
an absorbed disk blackbody model were tried.  If the fit was improved,
we accepted the better of these two alternatives as the correct model
for the source.  If an acceptable fit still was not found, a neutron
star atmosphere (NSA) model was tried.  If the NSA model had a better
overall fit or a similar fit with a more reasonable absorption
correction than the other 3 models, we accepted it as the correct
model for the source.  The best-fitting models were then used to
determine the absorption-corrected 0.3--7 keV luminosities for the
lightcurves.  Examples of power-law and blackbody fits are shown in
Figure~\ref{spec.examples}.  All but a few of our final fits were
acceptable ($\chi^2/\nu < 1.5$), and they provided the spectral
parameters and fit statistics shown in Table~\ref{spec.dat}.

In cases with high count rates, we tried adding a pileup model
\citep{davis2001,davis2003} to the model spectrum.  In these cases, we
included ``afterglow'' events, as is the standard when applying this
model.  All other aspects of the fitting procedure remained the same.
If the fit including pileup was an improvement according to the
$F$-test, we included pileup in our final determination of the model
spectral parameters.  Source r2-67 was the only one bright enough to
require a pileup model.  Our fits to the 2001 August 31, 2001 October
05, 2001 November 19, and 2001 December 07 observations of r2-67
employed pileup models with pileup fractions of 0.53, 0.36, 0.27, and
0.22 respectively.

The absorption-corrected 0.3--7 keV luminosities from the spectral
fits are provided in Table~\ref{spec.dat}, assuming a distance of 780
kpc.  If a spectral fit was performed on the source during any epoch,
conversions from flux to luminosity for all low-count detections, HRC
detections, and upper-limits were performed using PIMMS version
3.4a,\footnote{http://heasarc.gsfc.nasa.gov/docs/software/tools/pimms.html}
assuming the spectral parameters from the most contemporaneous
spectral fit and the effective areas as of 2003.  If no spectral fit
was available for the source in any epoch (s1-79, s1-82, r1-28, r2-69,
r2-72, r3-46, n1-59, n1-85, and n1-88) conversions from flux to
luminosity for all detections and upper-limits were performed using
PIMMS, assuming a power-law spectrum with photon index 1.7, $\rm
N_H$=10$^{21}$ \cm-2, and a distance of 780 kpc.

\begin{figure}
\centerline{\psfig{figure=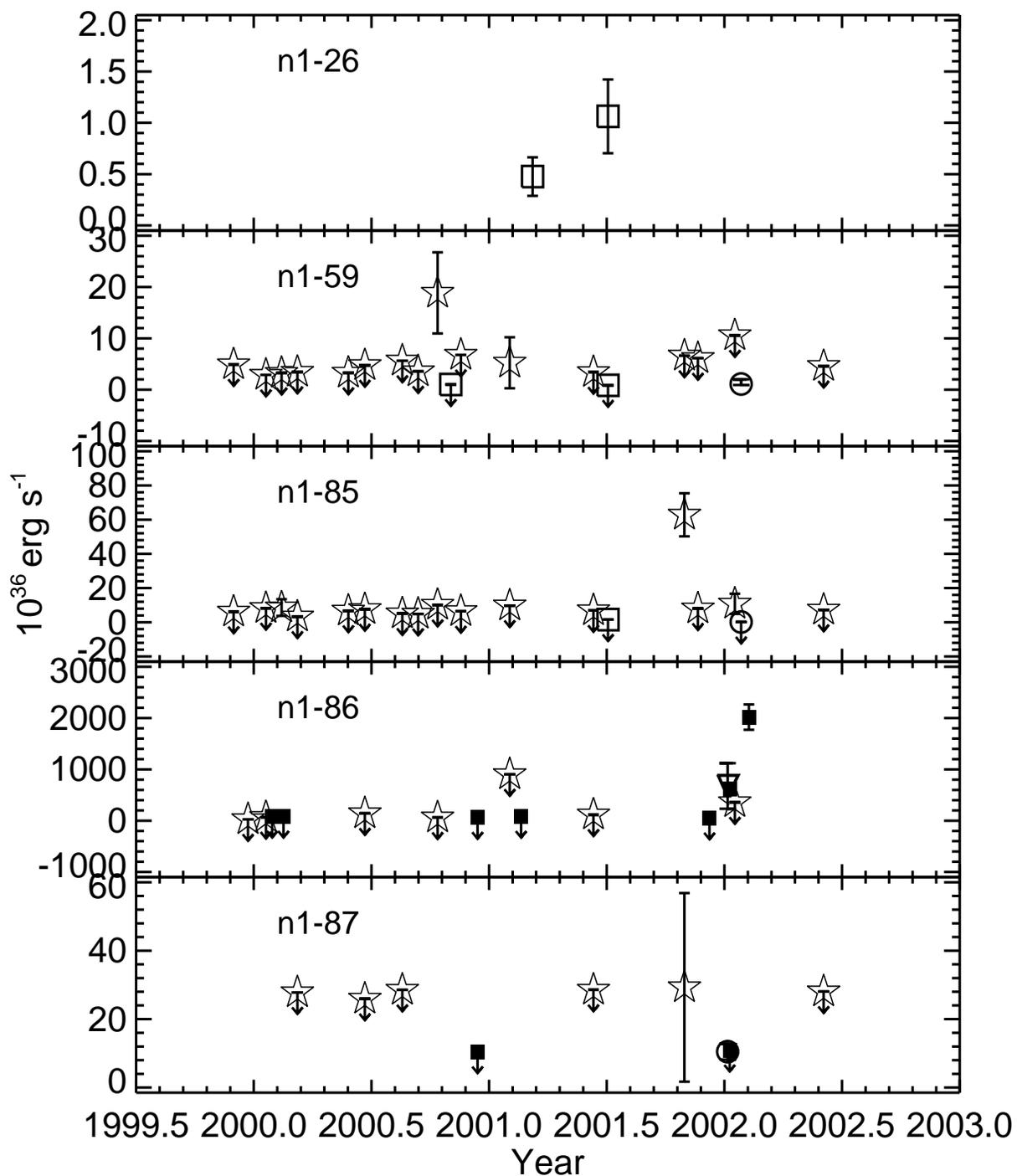,height=8.0in,angle=0}}
\caption{The lightcurves of the transient sources.  These lightcurves
contain data from XMM (circles, triangles mark cases where no PN data
was available), ACIS-I (filled squares), ACIS-S (open squares), and
HRC (open stars).  Upper-limits of non-detections are indicated with
arrows.}
\label{lc1}
\end{figure}

Finally, all of the luminosities and upper-limits were combined into
long-term lightcurves for each transient source.  These lightcurves
are shown in Figures~\ref{lc1}--\ref{lc9} and discussed in
\S~\ref{lightcurves}.

\subsubsection{Poor Spectral Fits}

There were objects with detections that contained $>$50 counts for
which it proved difficult to obtain a reasonable spectral fit using
the usual models.  The SSSs were especially difficult to fit (see
\citealp{distefano2004} for details about these sources).  Therefore,
the luminosities provided in Table~\ref{spec.dat} for these sources are
less reliable than for the non-SSS.  For example, the best fit for the
SSS r2-61 had $\chi^2_{\nu} = 3.7$.  

There was no single model type that fit every source satisfactorily.
While many sources were well-fitted by an absorbed power-law model,
several required other model types and/or multiple components to
produce an acceptable fit.  The worst case of this difficult fitting
was r2-67.

Two spectra of the transient source r2-67 were studied in detail in
\citet{williams2004}, and pileup was found to be important while the
source was bright.  We therefore included pileup in our fit to the
detection of r2-67 in OBSID 1575.  With 21,000 counts (0.6 ct
s$^{-1}$), this is the highest number of counts that we have for any
single {\it Chandra} ACIS observation of a transient source in M31,
and we were therefore able to compare several different model
combinations, including combining power-law with disk blackbody and
power-law with single-temperature blackbody, all with absorption.
After fixing the absorption column to $1.2\times 10^{21}$ cm$^{-2}$,
the best absorption from the {\it XMM-Newton} data and the January
2002 observation (ObsID 2897), a power-law fit with pileup yields a
spectral index of 2.2 and $\chi^2_{\nu} = 10.9$ for 125 degrees of
freedom ($dof$).  A disk blackbody model with pileup yields an inner
disk temperature of 0.62 keV and $\chi^2_{\nu} = 4.8$ for 125 $dof$, and
a single-temperature blackbody model with pileup did not converge.

After attempting these combination models, along with single component
models (no pileup) of all three types, we were unable to find a
$\chi^2_{\nu}$ fit better than 1.5, which is quite poor considering
the large number of degrees of freedom in this data set.

The only fit better than $\chi^2_{\nu}=2$ with fixed absorption was an
absorbed power-law with a disk blackbody component including pileup.
This model had $T_{inn} = 0.48\pm0.01$ keV, $\Gamma = 2.9\pm0.3$, and
a pileup fraction of 36\% ($\chi^2_{\nu} = 1.62$ with 122 $dof$), with
a unabsorbed 0.3--7 keV luminosity of 3.2$\times 10^{38}$ \ergs and
84\% of the emission is from the disk blackbody component.  The
$\chi^2_{\nu}$ value does not improve if absorption is a free
parameter.  The pileup component has a dramatic effect. The absorbed
power-law with a disk blackbody component without pileup was much
worse than including pileup ($\chi^2_{\nu} = 2.81$ with 120 $dof$).
If we allow absorption to be a free parameter, the fit improves
somewhat ($\chi^2_{\nu} = 1.84$ with 120 $dof$), but is still not as
good as the fit including pileup.  With the high counting rate and the
obvious readout streak in this data set, pileup is clearly a concern.
In addition, an $F$-test shows that there is $<$1\% probability that
pileup is not important.  Therefore, we used the disk blackbody plus
power-law with fixed absorption and pileup to estimate the luminosity
for this detection of r2-67.  The fit and residuals are shown in
Figure~\ref{r267}.

\subsection{XMM Data}\label{xmm}

Following the detection of X-ray transients in M31 with {\it Chandra},
we analyzed all of the publicly available {\it XMM-Newton} data
containing the M31 disk.  All isolated transient events outside of the
crowded 2$'$ X 2$'$ region surrounding the nucleus were analyzed.  The
raw events were processed and filtered using the updated calibration
database and analysis software package (SAS v5.4.1).  The energy
spectra of X-ray transients in M31 were extracted using the SAS
program {\it xmmselect} by selecting circular regions of radius in the
range of 10--15 arcsec centered at the source positions.  The
background spectra were extracted from the nearby source free
regions. The EPIC corresponding responses and effective areas were
generated by using the SAS tasks {\it rmfgen} and {\it arfgen},
respectively.  The spectra were binned to a minimum of 20 counts per
bin; for faint sources the spectra were binned to a minimum of 10
counts per bin. The spectral fittings were performed with the XSPEC
package. Simultaneous spectral fitting was performed when spectra for
more than one detector were available.  The best-fit model parameters
are given in Table~\ref{spec.dat}.

Upper-limits were measured at the positions of transient sources
detected in other observations.  The 4$\sigma$ detection limit at the
transient position was calculated from the exposure and background
level at the location where the transient had been observed in other
observations.  The upper-limit calculations were performed by applying
the flux upper-limit to the model spectrum from the most
contemporaneous spectral analysis.  For sources where no useful
spectral data were available from any epoch (n1-26, n1-85, n1-88,
r1-28, r2-62, r2-65, r2-66, r2-69, r2-72, r3-46, s1-27, s1-69, s1-79,
s1-82, s2-62), a power-law spectrum with a photon index $\Gamma = 1.7$
and $\rm N_H$=10$^{21}$ \cm-2 was assumed as the typical spectrum, as
determined in \citet{shirey2001}.  This assumption is also reasonable
considering the measured spectral properties of the transient sample
(see Figure~\ref{spec.hist}).

\subsection{Blending}\label{blending}

Four of our transient sources, r1-9, r1-11, r2-3, and r2-62 are within
6$''$ of a known, persistent X-ray source.  The {\it XMM-Newton}
observations of these events were not reliable for determining fluxes
as they were contaminated by the neighboring persistent sources.  The
lightcurves for these objects therefore were limited to the {\it
Chandra} observations alone.

Source r1-9 is sandwiched between persistent sources 1$''$ to its
north and 2$''$ to its south.  The northern source is close to the M31
nucleus, but is unlikely associated with it \citep{garcia2001}.
\citet{distefano2001} have suggested that this source may be the hot
stellar core of a tidally disrupted giant star.

We separated the r1-9 lightcurve from those of the neighboring sources
in the HRC data by taking the lightcurve measured with a 0.7$''$ box
described in \citet{williams2004}.  Separating the source in the ACIS
data required reprocessing the event lists to remove the pixel
position randomization that is part of the standard ACIS data
reduction pipeline.  In addition, we applied the subpixel event
repositioning algorithm of \citet{li2004} in order to obtain the best
possible resolution in the nuclear region.  These techniques allowed
us to resolve r1-9 from the nearby sources in the ACIS observations,
so that we were able to perform spectral fits on many detections and
obtain upper-limits when the source was not detected.

Source r1-11 is only 3$''$ away from r1-12, a bright, persistent,
variable source in the central M31 bulge.  Object r1-11 did not
fulfill the transient criteria in the {\it Chandra}-HRC synoptic
survey \citep{williams2004}, possibly due to contamination from r1-12
during its weak epochs.  

We remeasured the HRC lightcurve of this source with updated CIAO
software, using a 3$''$ X 3$''$ box aperture, much smaller than the
8$''$ X 8$''$ box used by \citet{williams2004}, to avoid
contamination from r1-12.  In addition, we chose our extraction
regions for the {\it Chandra} spectra of r1-11 so that r1-12 was not
included, and we excluded observations where the two objects were not
well resolved (OBSIDs 1577, 2895, 2897, and 2898) because they were too far
off-axis.

\begin{figure}
\centerline{\psfig{figure=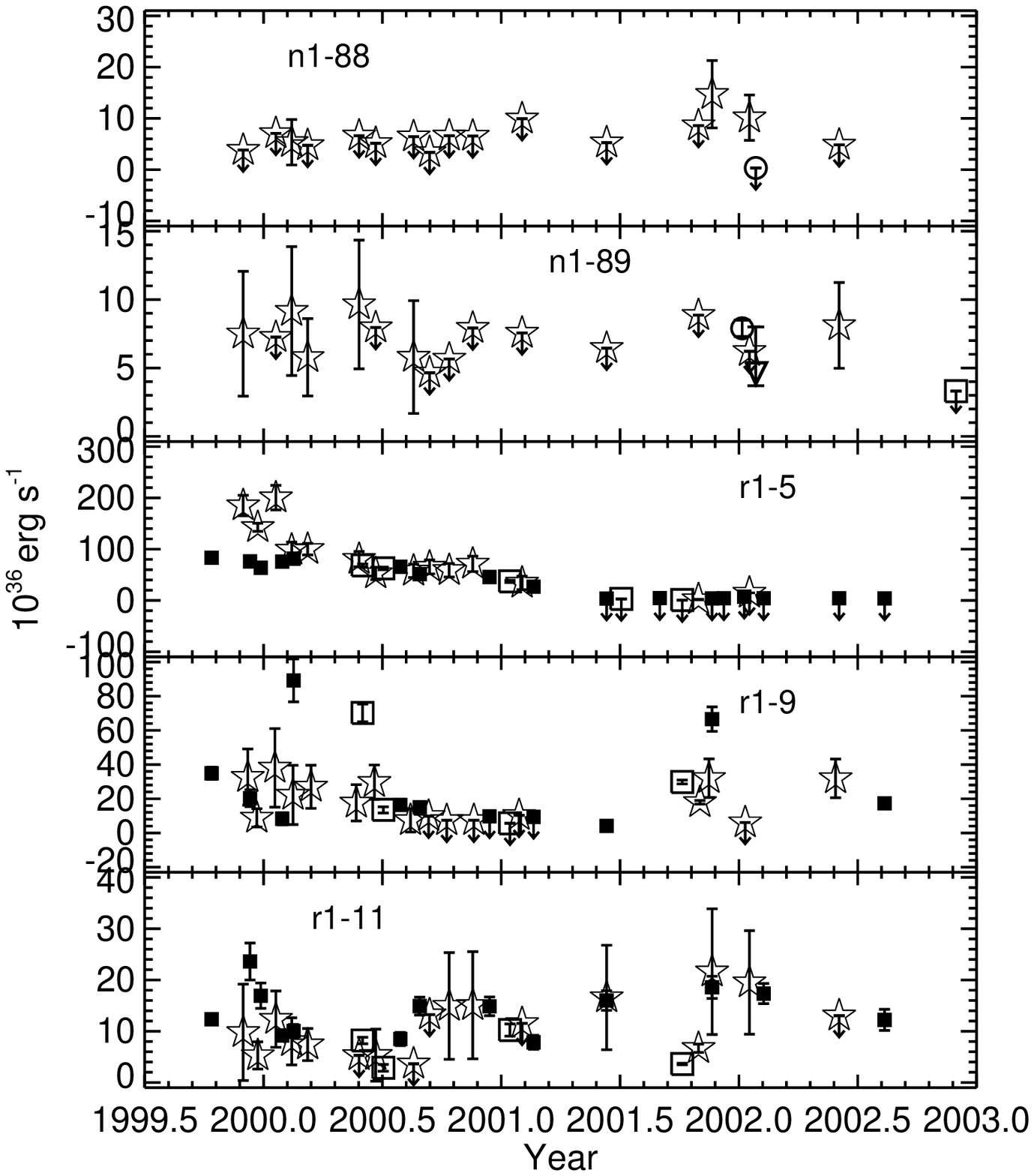,height=8.0in,angle=0}}
\caption{The lightcurves of the transient sources.  Symbols are the
same as Figure~\ref{lc1}.}
\label{lc2}
\end{figure}

\begin{figure}
\centerline{\psfig{figure=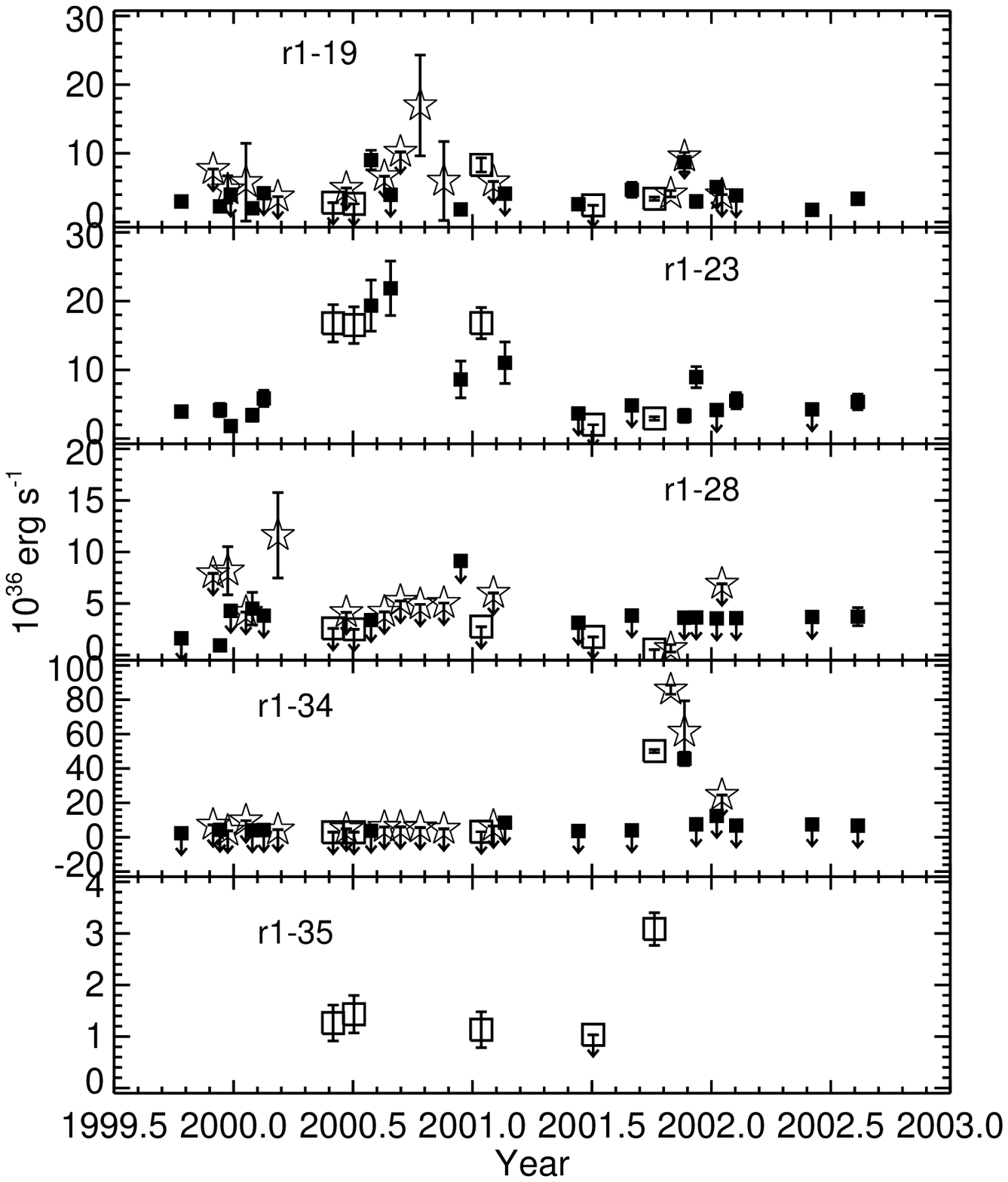,height=8.0in,angle=0}}
\caption{The lightcurves of the transient sources.  Symbols are the
same as Figure~\ref{lc1}.}
\label{lc3}
\end{figure}

\begin{figure}
\centerline{\psfig{figure=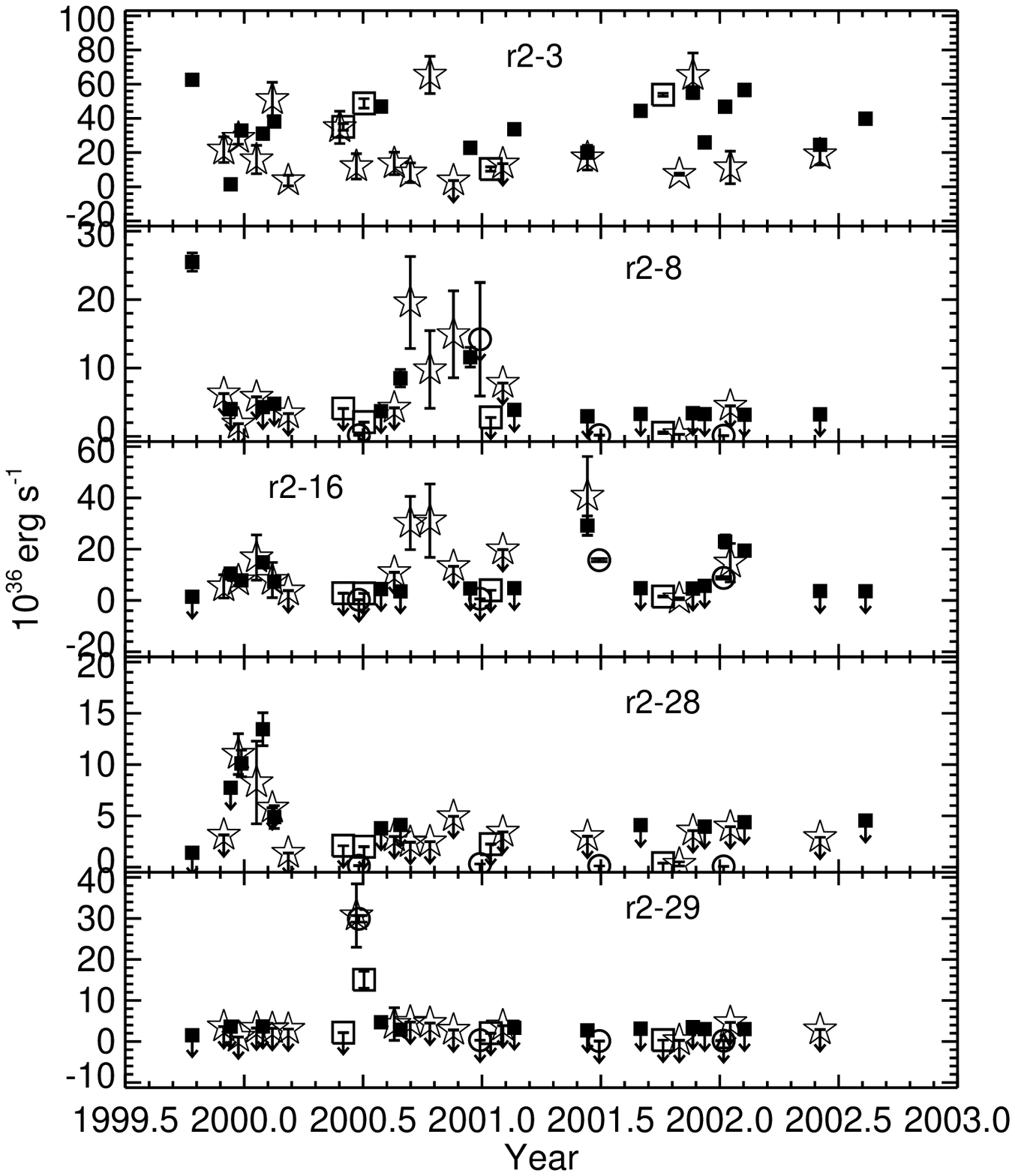,height=8.0in,angle=0}}
\caption{The lightcurves of the transient sources.  Symbols are the
same as Figure~\ref{lc1}.}
\label{lc4}
\end{figure}

\begin{figure}
\centerline{\psfig{figure=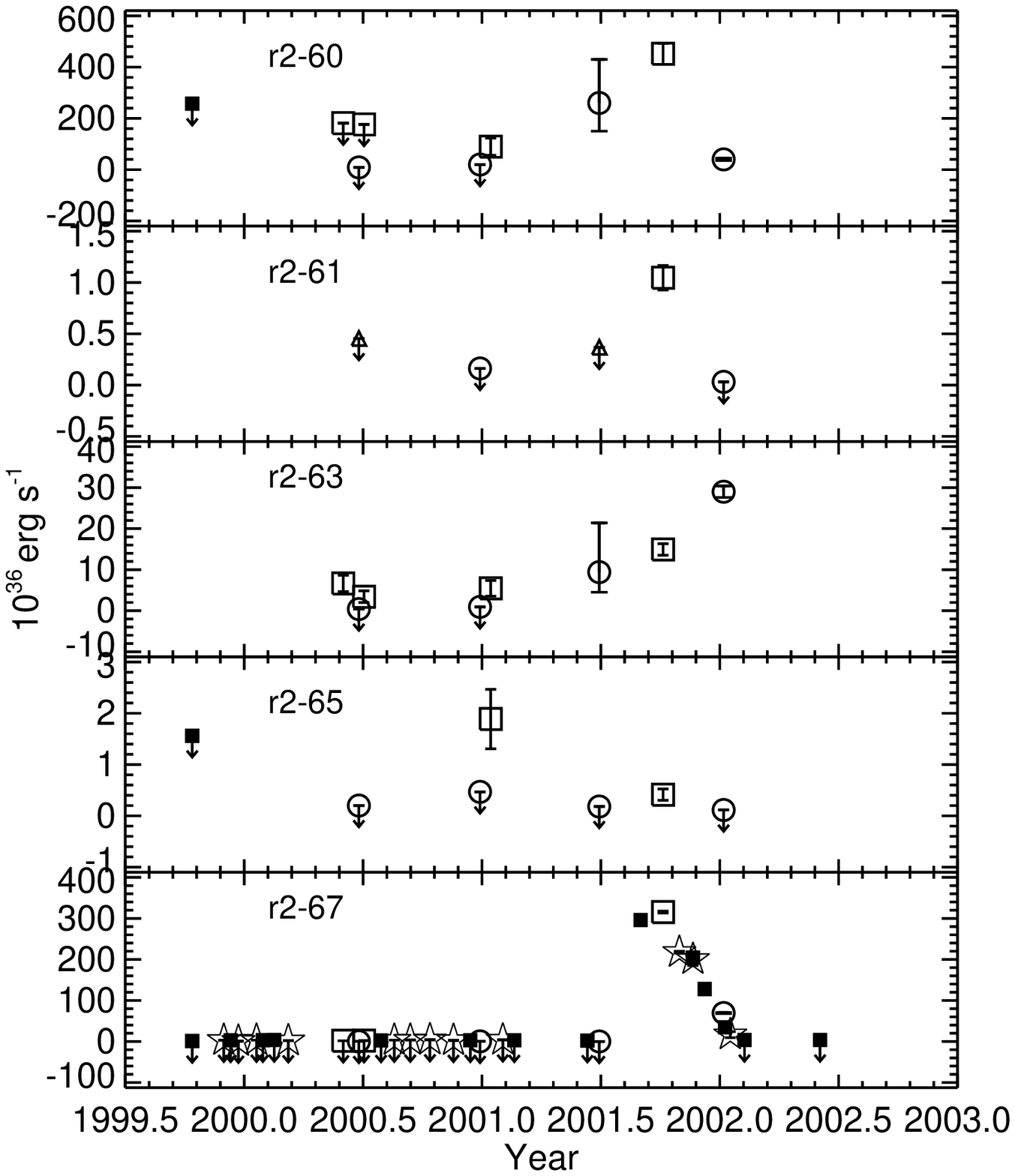,height=8.0in,angle=0}}
\caption{The lightcurves of the transient sources.  Symbols are the
same as Figure~\ref{lc1}.}
\label{lc5}
\end{figure}

\begin{figure}
\centerline{\psfig{figure=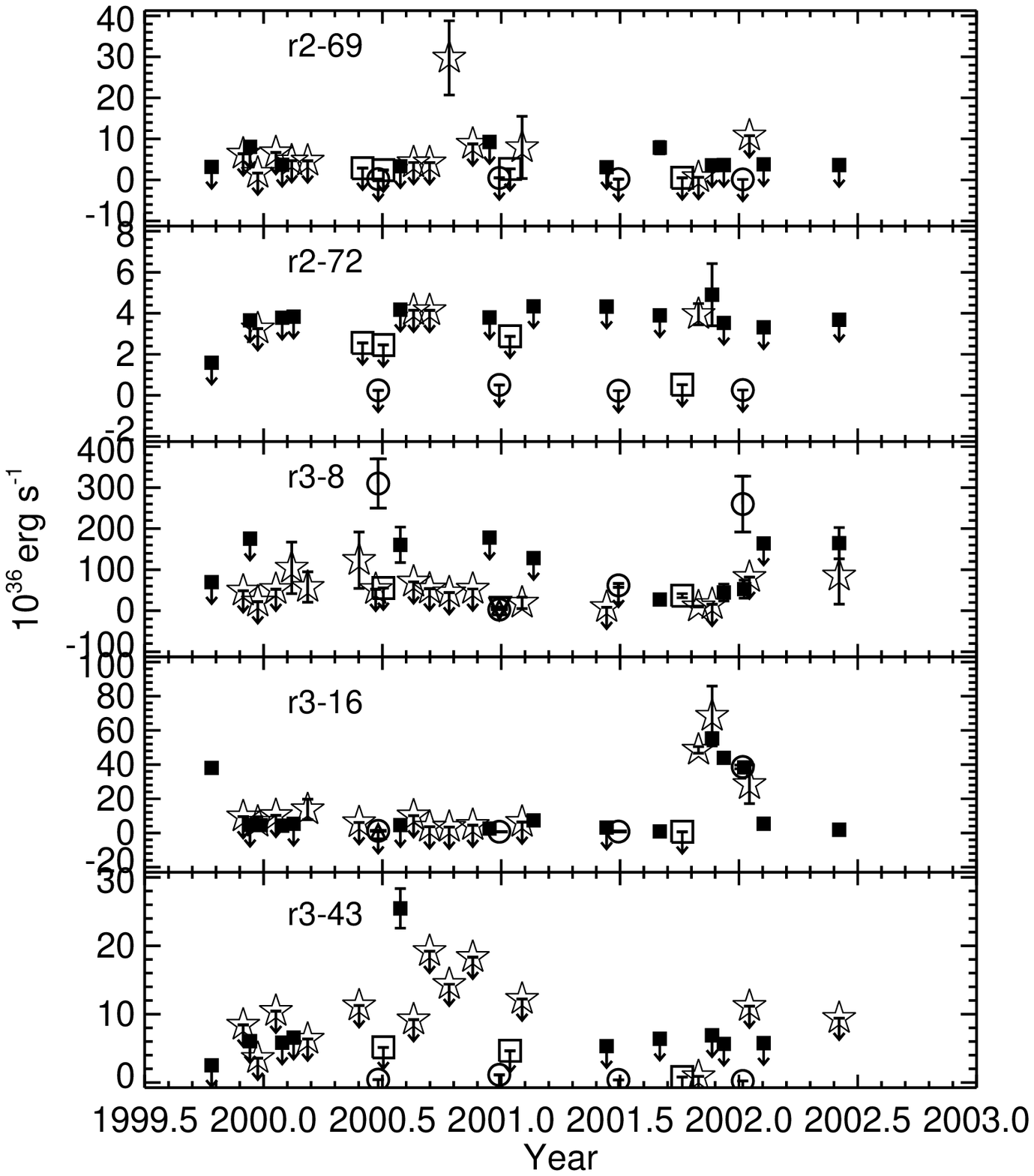,height=8.0in,angle=0}}
\caption{The lightcurves of the transient sources.  Symbols are the
same as Figure~\ref{lc1}.}
\label{lc6}
\end{figure}

\begin{figure}
\centerline{\psfig{figure=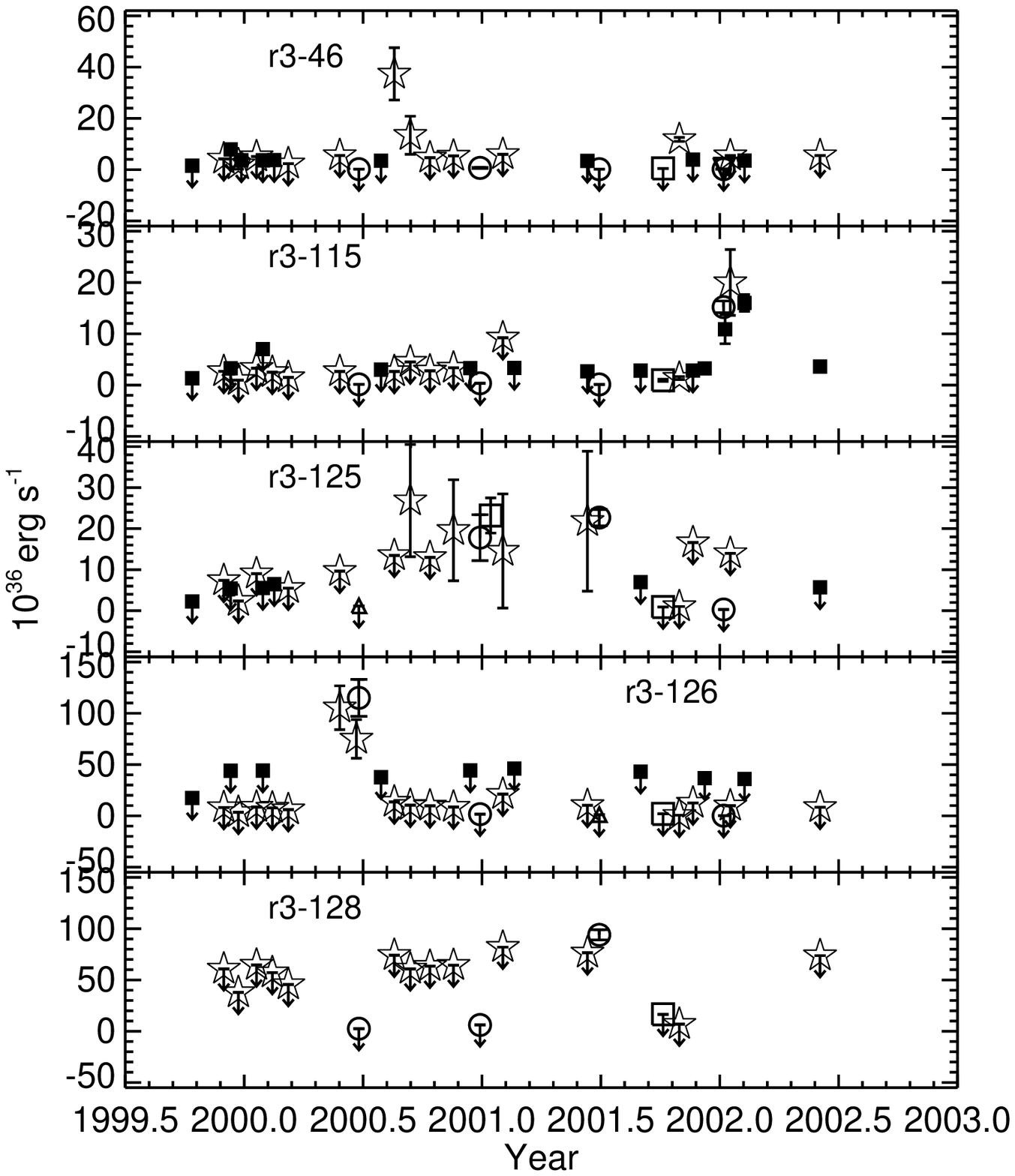,height=8.0in,angle=0}}
\caption{The lightcurves of the transient sources.  Symbols are the
same as Figure~\ref{lc1}.}
\label{lc7}
\end{figure}

\begin{figure}
\centerline{\psfig{figure=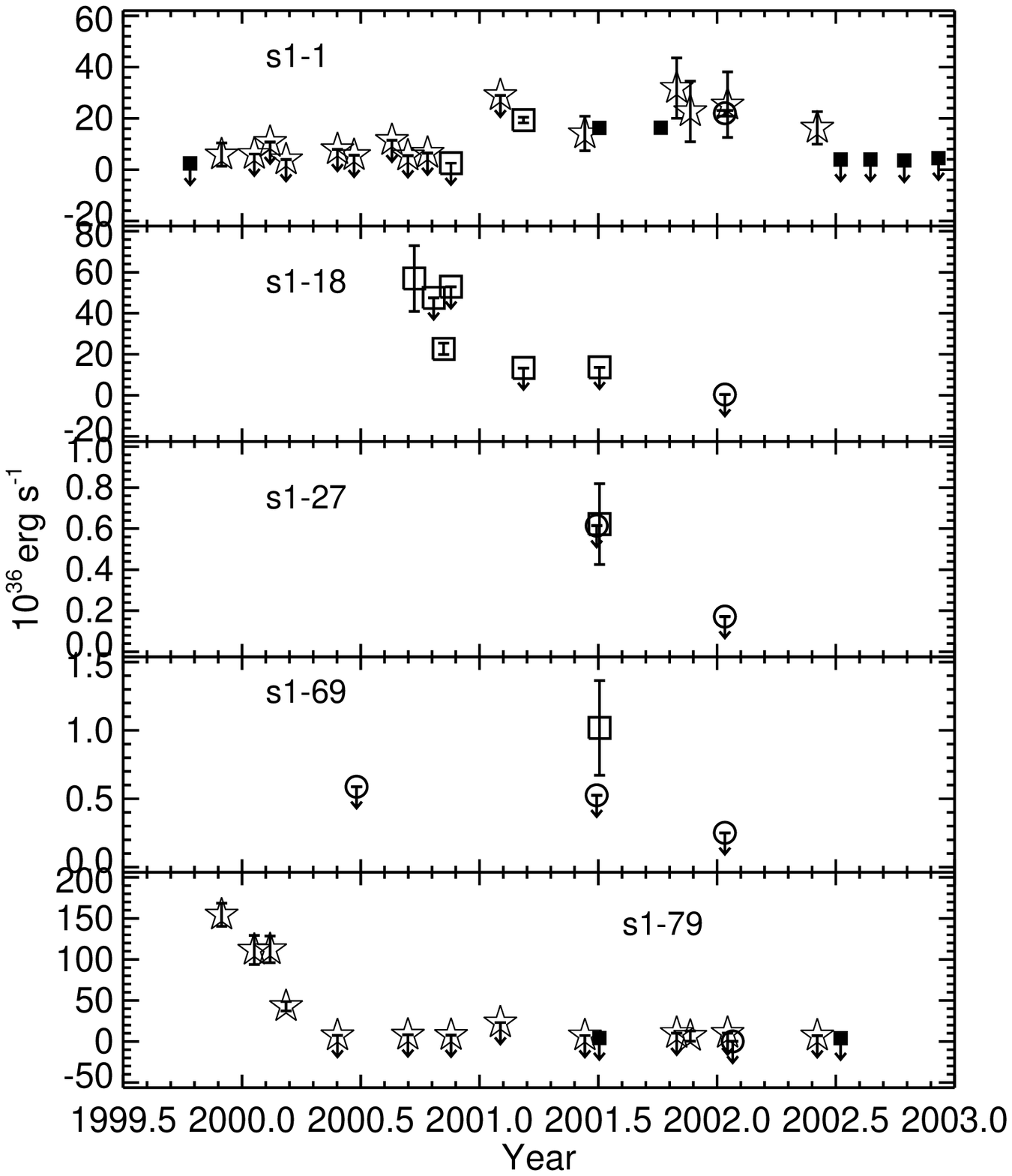,height=8.0in,angle=0}}
\caption{The lightcurves of the transient sources.  Symbols are the
same as Figure~\ref{lc1}.}
\label{lc8}
\end{figure}

\begin{figure}
\centerline{\psfig{figure=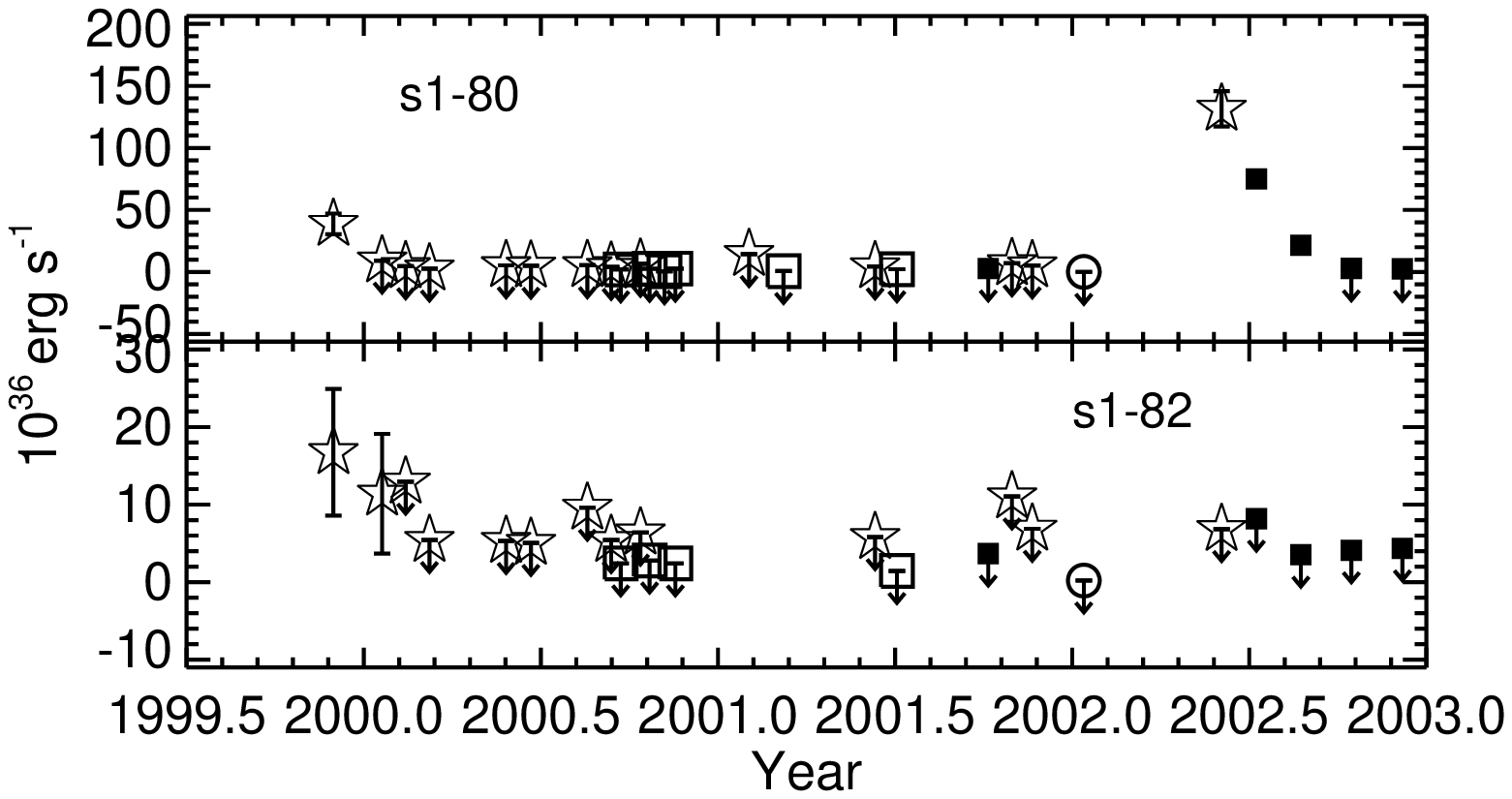,height=4.0in,angle=0}}
\caption{The lightcurves of the transient sources.  Symbols are the
same as Figure~\ref{lc1}.}
\label{lc9}
\end{figure}

\begin{figure}
\centerline{\psfig{figure=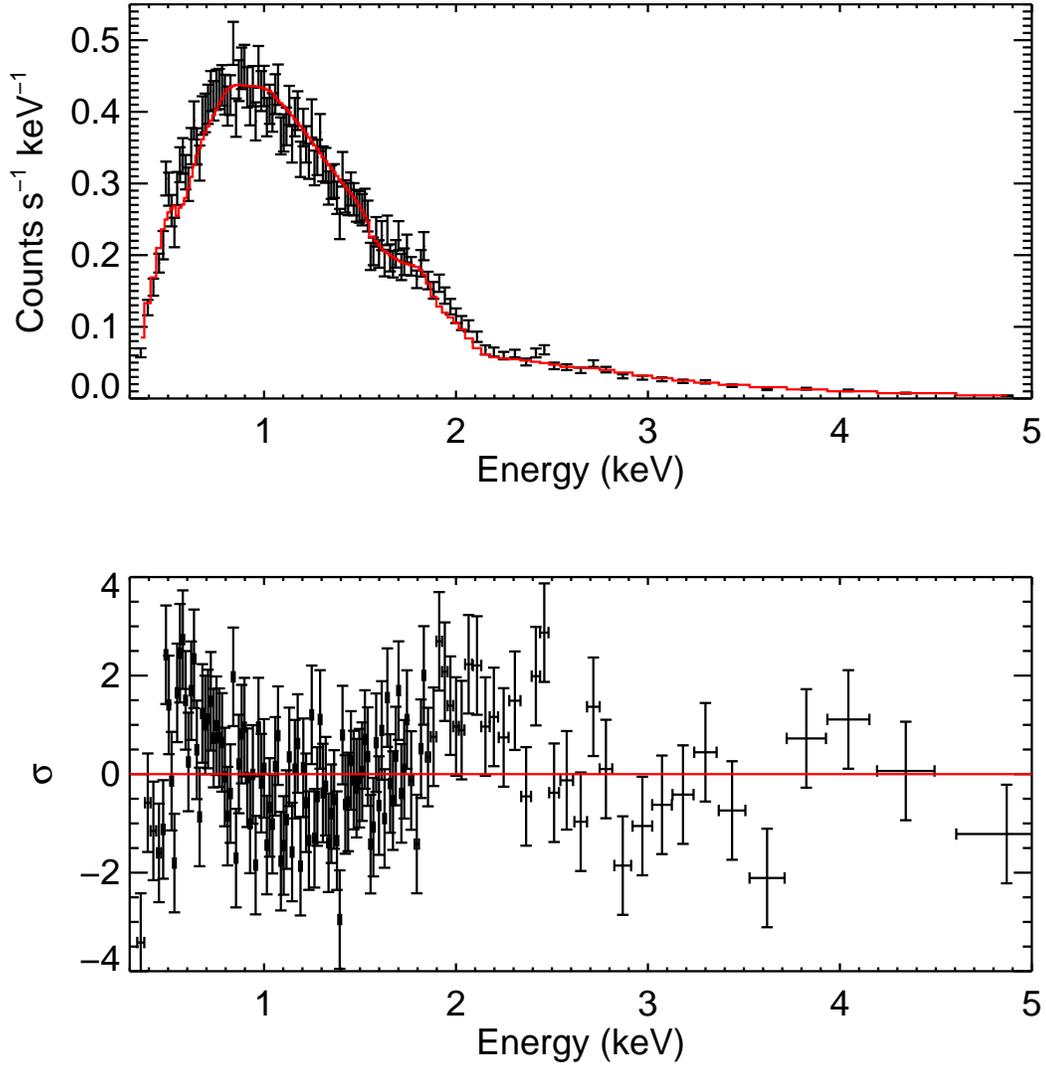,width=6in,angle=0}}
\caption{The best-fitting model (disk blackbody plus power-law) with
pileup for the 05-Oct-2001 (ObsID 1575) observation of r2-67. {\it
Top:} Histogram marks the best fit model prediction and error bars
mark the measured count rates from the data. {\it Bottom:} Error bars
mark the residuals between the model and the data in units of
$\sigma$.}
\label{r267}
\end{figure}

\begin{figure}
\centerline{\psfig{figure=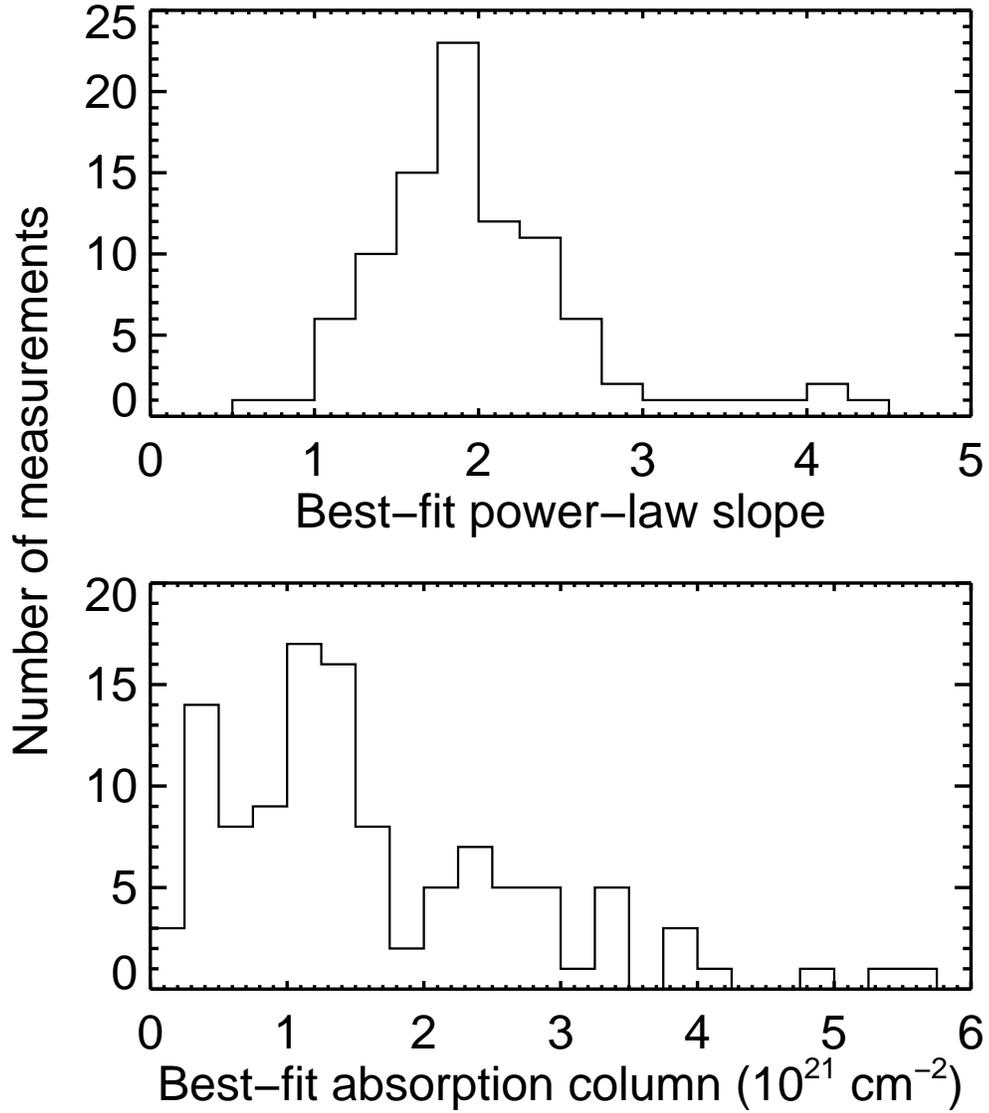,width=6in,angle=0}}
\caption{{\it Top:} Histogram of the power-law slope ($\Gamma$)
measurements for all detections fitted with power-law spectral models.
{\it Bottom:} Histogram of the absorption column measurements for all
spectral fits where the absorption was a free parameter and was
constrained beyond an upper-limit value.}
\label{spec.hist}
\end{figure}

This scrutiny of each data point ensures us that the lightcurve shown
in Figure~\ref{lc2} is free from contamination.  While this lightcurve
clearly shows variability at about the factor of 5 level, all
observations sensitive to better than 5 $\times 10^{36}$ \ergs\ detect
this source.  While there are some non-detections in the shallower HRC
data, the available data are consistent with this object being a
variable source with occasional dips below $\sim$5 $\times 10^{36}$
\ergs .  This variability would not fit the \citet{williams2004}
criteria for a transient, but we include the source here as it has
been classified as a transient source by \citet{kong2002}.

Source r2-3 is only 6$''$ (23 pc) from r2-4, a bright variable X-ray
source in the globular cluster Bol 146.  The proximity to Bol~146
suggests r2-3 could be a member of the cluster.  The cluster light has
a full-width half-maximum of 1.2$''$ in the images of the Local Group
Survey \cite{massey2001}, indicating a core radius of $\sim$2.3 pc and
placing r2-3 $\sim$10 core radii from the cluster center.  If r2-3 is
a cluster member, it is the only transient in our catalog associated
with a globular cluster.  Object r2-3 did not fulfill the transient
criteria in the {\it Chandra}-HRC synoptic survey
\citep{williams2004}, possibly due to contamination from r2-4 during
its weak epochs.

We remeasured the HRC lightcurve of this source with updated CIAO
software, using a 4$''$ X 4$''$ box aperture, much smaller than the
8$''$ X 8$''$ box used by \citet{williams2004}, to avoid contamination
from r2-4.  In addition, we chose our extraction regions for the {\it
Chandra} spectra of r2-3 so that r2-4 was not included.  Because of
the separation distance, this was possible for all observations.

In contrast to r1-11, the resolved lightcurve of r2-3 appears
extremely variable, with near factors of 100 changes in luminosity on
timescales of months.  This source may in fact be a good transient
candidate; however, if it is a BHB, it has a very high duty cycle.  It
is brighter than 10$^{36}$ \ergs\ in all but 2 observations, and these
observations both have upper-limits $>$10$^{36}$ \ergs .

Source r2-62 is 5$''$ from the persistent, variable source r2-25
\citep{kong2002}. The source was only detected by the deepest {\it
Chandra} ACIS-S observation.  Since this lightcurve contained only one
relevant data point (all upper-limits were above the one detected
luminosity), we did not include the lightcurve for this source in
Figures~\ref{lc1}-\ref{lc9}, nor did we attempt to estimate its
properties in Table~\ref{etime}.

\section{Results}\label{results}

\subsection{Location}\label{locationsec}

We cross-correlated our sample of transients against catalogs of
globular clusters in M31 and found no clear associations.  Although
r2-3 could be associated with Bol~146 (see \S~\ref{blending}), the
6$''$ separation makes cluster membership unclear.  Even so, the
location of the transient sources can reveal something about their
age.  As discussed in \citet{williams2004}, transient events in the
bulge are more likely to be black hole X-ray novae because they are
likely old and therefore unlikely to be high-mass X-ray binaries.
Studies of transient Galactic LMXBs show a large fraction of them to
contain black holes.  Therefore it is especially interesting to see
the apparent clustering of transient events to the central bulge of
M31.

\begin{figure}
\centerline{\psfig{figure=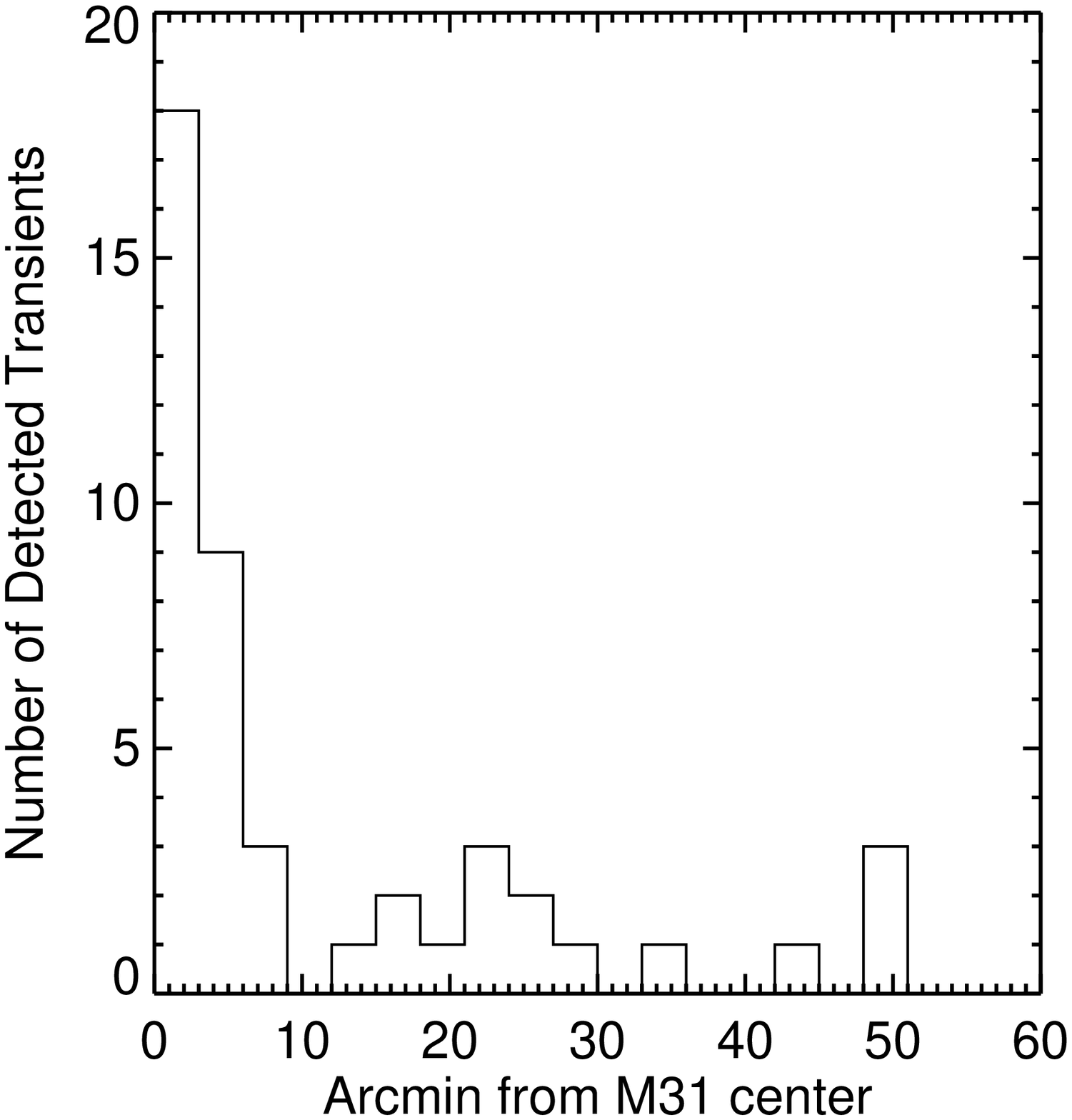,width=6in,angle=0}}
\caption{A histogram of the number of detected transients as a
function of their distance from the center of M31 is shown.  The
distribution shows a clear peak near the center of the galaxy,
suggesting most transient X-ray sources in M31 are old objects.}
\label{location}
\end{figure}

Figure~\ref{location} shows the distribution of detected transient
sources as a function of projected distance from the center of M31.
Clearly most of the events lie within ~10$'$ (2.3 kpc) of the center
of the galaxy.  In fact, 30 of 45 events lie with 8$'$ (1.8 kpc) of
the center of the galaxy, and 39 of 45 lie within 0.5$^{\circ}$ (6.8
kpc) of the center.

Clearly some of this central concentration is due to the fact that the
center of M31 has been observed far more frequently than the rest of
the galaxy; however, even the unbiased {\it Chandra}-HRC survey of
\citet{williams2004} found half of the transient events detected
occurred in the bulge.  In addition, there is a clear gradient within
the region of the galaxy most surveyed.  With fields of view of 17$'$
$\times$ 17$'$ and 30$'$ $\times$ 30$'$, typical {\it Chandra} and
XMM-{\it Newton} observations that are centered within 2$'$ of the
nucleus are unbiased out to 8$'$, yet there is a clear decrease in the
number of detected events with radius from the galaxy center to 8$'$
away.  Even though the {\it Chandra} PSF expands to a size of 5$''$ at
8$'$ off-axis, because of the low background and low source density
outside of the central 2$'$, the sensitivity of our typical 5 ks ACIS
observation is only $\sim$10\% lower 8$'$ off-axis than on-axis.
Therefore, while there have likely been more events in the disk that
have gone unobserved than events in the bulge that have gone
unobserved, the events still appear to be clustered toward the galaxy
center.

As the M31 bulge contains little or no recent star formation, these
X-ray binaries are most likely old ($\gap$1 Gyr).  Therefore most of
them are likely LMXBs.  Since a high percentage of Galactic LMXBs that
produce bright X-ray novae contain dynamically confirmed black hole
primaries, the locations of these events suggest that they arise
primarily from BH-containing X-ray binaries.
 
\subsection{Lightcurves}\label{lightcurves}

The lightcurves for all of the transient sources detected in at least
one observation are shown in Figures~\ref{lc1}--\ref{lc9}, including
all observations during the relevant time period made by {\it
XMM-Newton}, {\it Chandra} ACIS and {\it Chandra} HRC.  To limit
confusion, if there were 2 or more upper-limits measured within one
week, only the faintest upper-limit is plotted in the figure.  All
upper-limits brighter than the brightest detection were removed.
 
Some lightcurves have observations that appear very close in time but
yield different luminosity measurements.  The long time line can be
deceiving in this regard as these observations are separated by at
least several days.  These sources can change in luminosity by factors
of 10 or more in that time span. For example, in the lightcurve of
r1-5 the first three HRC detections appear to be significantly
brighter than the early ACIS detections.  Unfortunately, since none of
the observations were simultaneous and the HRC yields no spectral
information, we can only conclude that the source was either brighter
or had a different spectrum during those HRC observations. In cases
where the observations truly were within a day or two of simultaneous
(see Table~\ref{spec.dat} for dates), the luminosities from different
instruments are consistent within the errors.

Those sources without any bright detections (n1-26, r1-35, r2-61,
s1-27, s1-69) could not be classified using the available data.
Sources r2-66 and s2-62 were not detected in any single epoch
observation.  We therefore did not include lightcurves for these
sources at all, as they were only detected by stacking images from
multiple epochs in \citet{distefano2004}. Source r2-62 was only
detected in the deepest ACIS-S observation, and there were no
upper-limits for this source that showed it was fainter than this
luminosity.  We therefore did not include the lightcurve for r2-62
either.  These eight sources do not fit the criteria for an X-ray
transient used in \citet{williams2004}, but were classified as such by
the criteria of \citet{distefano2004}.

We fit an exponential decay to each lightcurve assuming the brightest
detection as the lightcurve peak and assuming half of the upper-limit
flux (and an error equal to half of the upper-limit) for HRC
(1$\sigma$) upper-limits, and one third of the upper-limit flux (and
an error equal to one third of the upper-limit) for ACIS and XMM
(4$\sigma$) upper-limits.  This method was reasonable because the ACIS
and XMM upper-limits were more conservative than those taken from the
much shallower HRC data.  The fit was terminated when there were 2
non-detections in succession.  Such exponential decay is
representative of the overall shape of the lightcurves of most
Galactic X-ray novae \citep{chen1997}.  Lightcurves for sources with
no bright detections (n1-26, r1-35, r2-66, s2-62) or a decay curve
with fewer than 3 data points between luminosity peaks (n1-26, n1-86)
could not provide a reliable estimate of $e$-folding time.  In some
cases (n1-89, r2-60, r2-61, s1-27, s1-69) there was only a single
non-detection after the brightest detection.  For these two-point
decay curves, if the upper-limit was at least a factor of $e$ fainter
than the detection, the decay time was constrained to less than the
time between the two observations.

These $e$-folding decay time estimates help to characterize the length
of the burst, but they are only estimates based on our coarsely
sampled lightcurves.  We note that the time separation of the data
points can be three or four months in many cases.  In addition, the
lightcurves may not have decayed exponentially in all cases
\citep{shahbaz1998,williams2005bh1}.  For these reasons, the decay
time measurements should only be considered rough approximations.

We counted the distinct outbursts in the lightcurve of each source
that showed such distinct outbursts.  For example, r2-67 clearly shows
only one outburst during the studied time period.  Source r2-8 shows 2
outbursts clearly separated by several months of non-detections.  This
number, given in Table~\ref{etime}, helps to show the burst frequency
of the sources.  The majority (23) of the XRTs showed only 1 clear
outburst.  Seven sources showed 2 outbursts, and 3 sources showed 4
outbursts.  The remaining 9 XRTs detected in single epochs were either
too faint (e.g. n1-26 and r1-35) or had lightcurves that were too
complex (e.g. r1-23 and r2-3) to clearly count outbursts.

\begin{figure}
\centerline{\psfig{figure=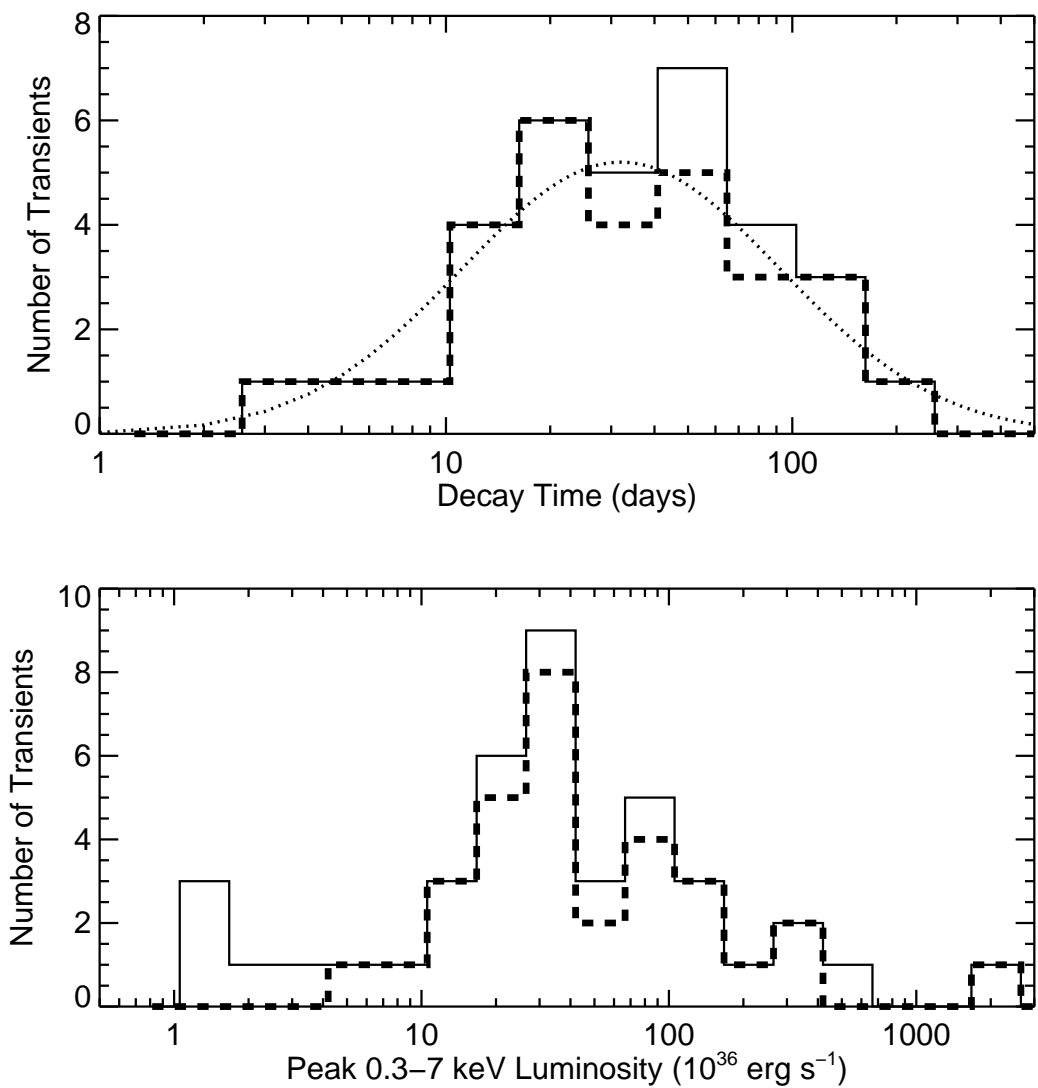,width=6in,angle=0}}
\caption{{\it Top:} A histogram of the number of transients as a
function of $e$-folding decay time. {\it Solid line} gives all
transients with measured decay times, and {\it dashed line} ignores
the SSSs and QSSs. {\it Dotted line} shows best-fitting Gaussian to
the distribution.  The peak and width of the Gaussian are consistent
with the distribution found for Galactic transients by
\citet{chen1997}. {\it Bottom:} A histogram of the number of
transients as a function of maximum observed luminosity.  {\it Solid
line} gives all transients with at least one single-epoch detection,
and {\it dashed line} ignores the SSSs and QSSs.  The single source
above 10$^{39}$ erg s$^{-1}$ is n1-86, a soft source that may be a
foreground CV, which would lower its peak luminosity to $<10^{36}$ erg
s$^{-1}$.}
\label{etimefig}
\end{figure}

The date and unabsorbed (0.3--7 keV) luminosity of the brightest
detection, the number of lightcurve points used in the fit, the
faintest upper-limit, the bright/faint luminosity ratio, the best-fit
$e$-folding decay time, the number of distinct outbursts observed, the
duty cycle estimate, and the lightcurve type for each transient source
are provided in Table~\ref{etime}.  In addition, a histogram of the
number of transients as a function of $e$-folding time is given in
Figure~\ref{etimefig}, showing a peak at $\sim$25 days for the
non-SSSs.

A similar peak was seen for Galactic outbursts \citep{chen1997}.  For
comparison, we fit our distribution with a Gaussian using
least-squares.  The best fit, shown with the dotted line in
Figure~\ref{etimefig}, has a $<log(\tau_d)> = 1.5\pm0.5$.  This
value is consistent with the Galactic distribution of $<log(\tau_d)> =
1.24\pm0.36$.  The SSSs and QSSs have longer decay times, but these
lightcurves are the most poorly constrained, with large correction
factors for their luminosity estimates and peak detections only a
factor of a few brighter than the upper-limits of the non-detections.

\begin{figure}
\centerline{\psfig{figure=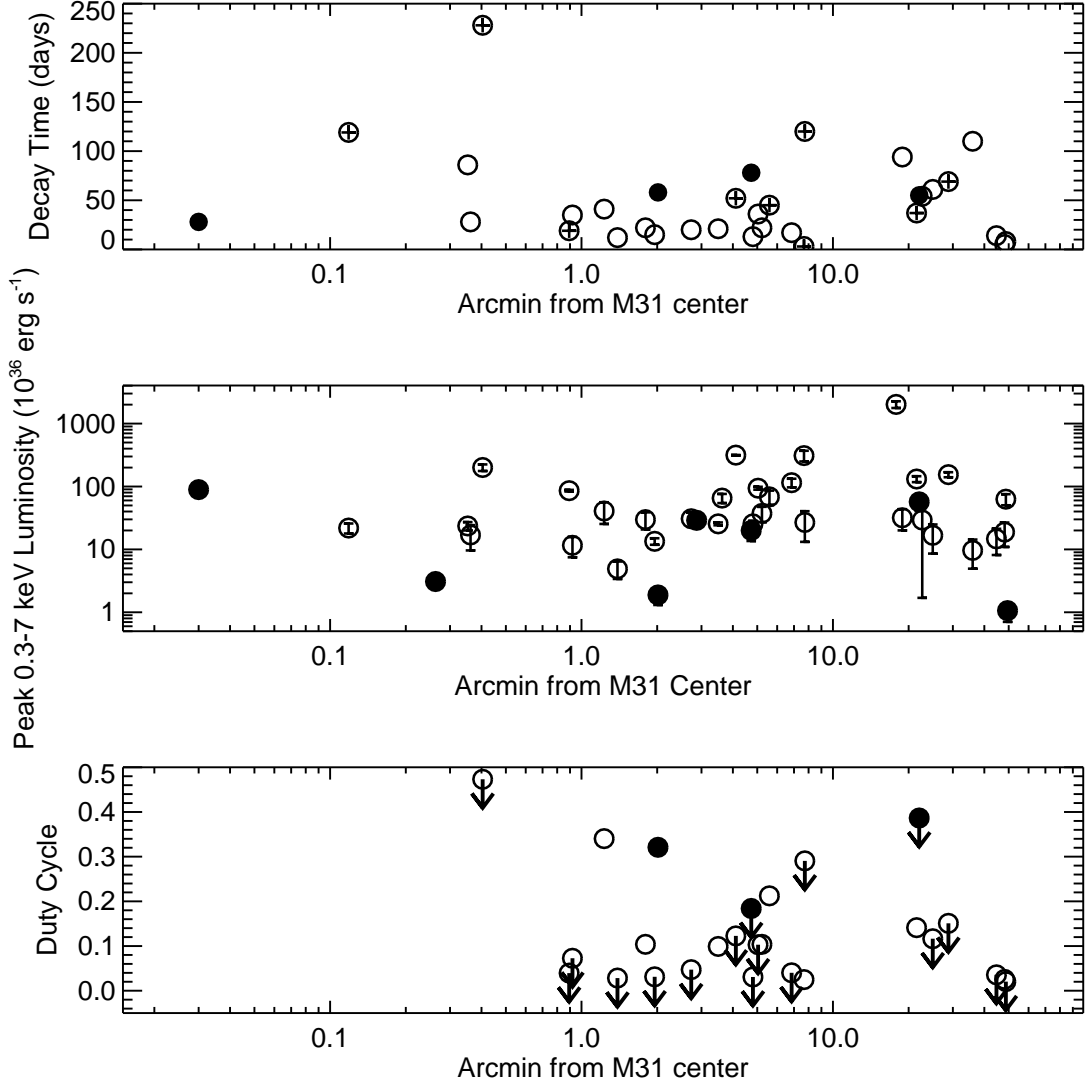,width=6in,angle=0}}
\caption{{\it Top:} A plot of the $e$-folding decay times of the XRTs
as a function of their distance from the galaxy center.  Filled
circles denote SSSs and QSSs.  Circles around crosses mark poor fits
to an exponential decay ($\chi^2_{\nu} > 3$).  {\it Middle:} The
maximum observed luminosities of the transients as a function of their
distance from the galaxy center.  Filled circles denote SSSs and
QSSs. {\it Bottom:} The duty cycle estimates of the transients as a
function of galactocentric distance.  Filled circles denote SSSs and
QSSs.  Upper-limits are plotted for sources showing only one outburst
during the time period studied.}
\label{efradfig}
\end{figure}

Figure~\ref{efradfig} shows the decay time estimates, maximum observed
luminosities, and duty cycle estimates of the XRTs as a function of
their projected distance from the galaxy center. No clear correlation
is found in the XRT properties with their position in the galaxy.  The
only noteworthy feature in the figures is the three faintest XRTs are
all SSSs (plotted with filled circles), hinting that these sources may
not be appropriate members of the XRT sample.  On the other hand,
three of the SSSs are indistinguishable from the rest of the sample in
both plots, indicating that they are appropriate members of the
sample.

\begin{figure}
\centerline{\psfig{figure=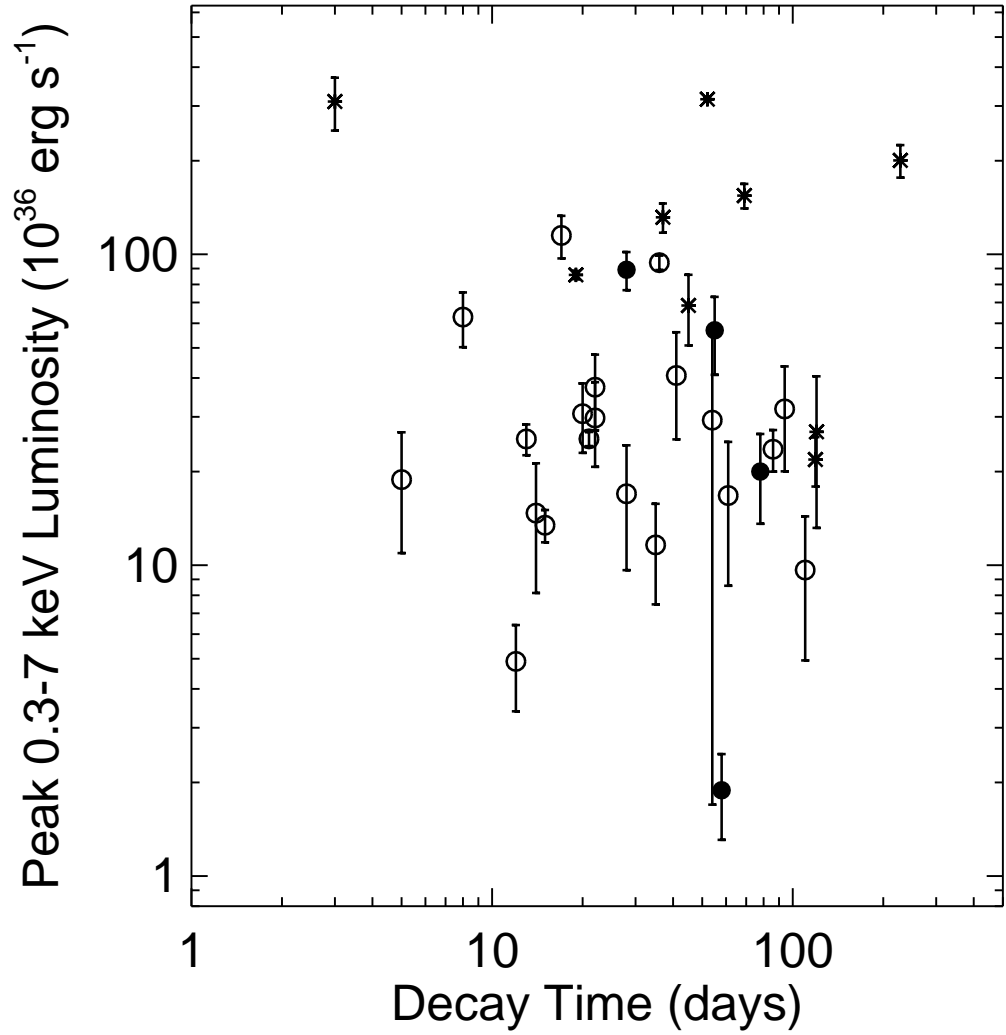,width=6in,angle=0}}
\caption{The $e$-folding decay time vs. the peak luminosity.  The
values cover a similar range in parameter space to those seen in
Galactic outbursts \citep{chen1997}.  Filled circles denote SSSs and
QSSs.  Asterisks mark poor fits to an exponential decay ($\chi^2_{\nu}
> 3$).}
\label{efl}
\end{figure}

A plot of the decay time vs. the peak luminosity of the transient
events is provided in Figure~\ref{efl}.  No correlation is apparent,
which adds confidence that there are no strong biases in the data or
analysis that would cause more luminous bursts to appear to have
shorter decays.  This figure also shows that the $\sim$30 day peak in
Figure~\ref{etimefig} contains sources covering 2 decades of peak
luminosity.  We note that our brightest luminosity transient, n1-86,
is not shown.  No decay time was measured for this source because its
brightest detection was in the final observation.

In addition, one identifier of accreting BHs is a luminosity
$>L_{edd}$ for a typical (1.4 $M_{\odot}$) NS.  Figure~\ref{efradfig} shows
a histogram of the peak observed luminosities of the M31
transients. Our sample contains 4 events brighter than
2.5$\times$10$^{38}$ erg s$^{-1}$, which fit this criterion.

In addition, the range of peak luminosities (see
Figure~\ref{etimefig}) is similar to the Galactic sample
\citep{chen1997}, considering the different sampling rates and number
of years of monitoring.  Their sample contains 33 sources, ours has 30
with measured decay times, not including SSSs or QSSs.  Their sample
has 12 events with peak luminosities $>$10$^{38}$ erg s$^{-1}$; ours
has 7.  Their sample has 3 events with peak luminosities $>$10$^{39}$
erg s$^{-1}$; ours has 1.  This source, n1-86, is very soft and has a
large absorption correction.  Furthermore, the effective temperature
of the spectral fit indicates that n1-86 may be a foreground source
(see \S~\ref{nsa}).  If so, its luminosity would be only
$\sim$10$^{35}$ erg s$^{-1}$.  

The lower number of bright peak luminosities may be due to the
sampling of our data and the length of our timeline.  Since our
lightcurves are only sampled at about once per 2 months, we are not
likely to have measured these transients at their absolute highest
luminosities, considering Galactic transients only stay at their peak
brightness for $\sim$days \citep{chen1997,mcclintock2003}.  In any case, the
\citet{chen1997} sample covers 26 years (1970--1996) of observations,
while our sample only covers 3 years (1999--2002).  If the rate of
10$^{39}$ erg s$^{-1}$ XRNe is $\sim$0.1 yr$^{-1}$, we should not
expect to see more than 1 in 3 years of monitoring M31.

\subsection{Duty Cycles}

The outburst numbers were used to estimate the duty cycles for XRTs
with measured decay times and peak/quiescent luminosity ratios
$>$15. These estimates are also provided in Table~\ref{etime}.  The
duty cycles for sources showing evidence for only a single outburst
are given as upper-limits, assuming that no outbursts were missed
during the course of the observations shown in the lightcurve.  It is
certainly possible that this assumption is incorrect for a few of the
more poorly sampled objects, but since most sources appear to be faint
more often then they are bright, we believe it more likely that they
were faint for significantly more time before and after our sampling
period than that we missed an outburst during the sampling period.
Therefore this assumption is more conservative than alternative
assumptions (e.g. that this measurement is a good estimate of the real
duty cycle or that this measurement is a lower limit to the real duty
cycle).  

\begin{figure}
\centerline{\psfig{figure=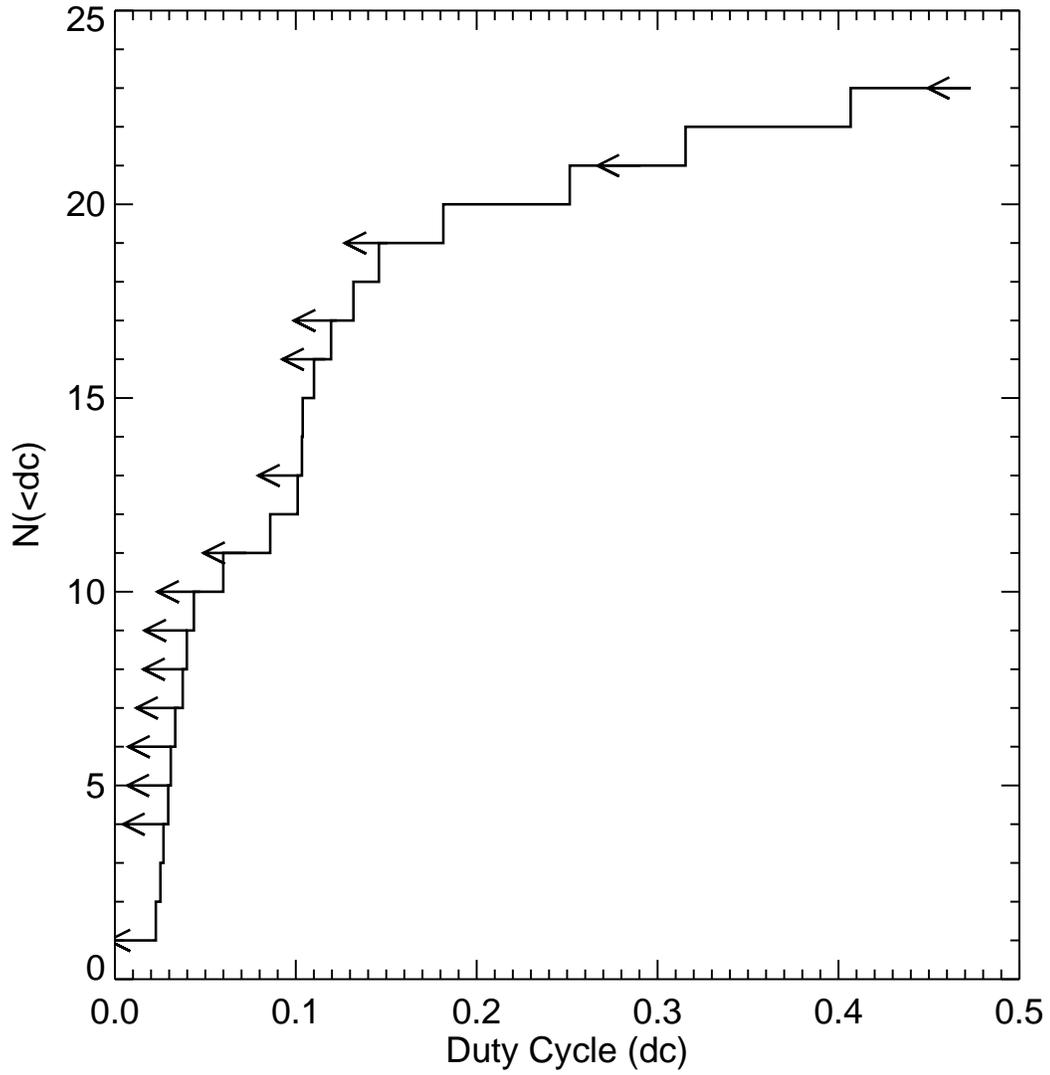,width=6in,angle=0}}
\caption{The distribution of duty cycles for the 23 non-SSSs with
constraints from the available data.  The histogram shows the
cumulative number of XRTs with duty cycles less than the values given
on the abscissa.  Upper-limit values are shown with left arrows.}
\label{dcfig}
\end{figure}

A histogram of the cumulative number of XRTs as a function of duty
cycle is shown in Figure~\ref{dcfig}.  About half of the measured duty
cycles are $<$0.1, and all but three are $<$0.2.  This distribution is
compatible with the Galactic sample of \citet{chen1997}, which shows
recurrence times ranging from months to $\sim$100 years, with most
recurrence times $>$1 year.  If one assumes the decay time of $\sim$20
days, this large range of recurrence times yields a wide range in duty
cycles from 0.0005--0.3, with most duty cycles $<$0.1.

The lightcurves reveal a wide variety of photometric behavior, but
they can be described as falling into four categories.  Category one
contains sources like r1-11, which are active during most epochs but
disappear occasionally.  We classify these as ``High Duty Cycle''
(HDC) transients.  Category two contains sources like r2-8, which
appear for a few months more than once over the course of the studied
time period.  We classify these as ``Medium Duty Cycle'' (MDC)
transients.  Category three contains sources like r1-5, which only
appear once but take years to disappear.  We classify these as ``Long
Decay'' (LD) transients.  Finally, category four contains sources like
r3-126, which appear only once for a short time, sometimes so short
that they are only detected in one observation.  We classify these as
``Low Duty Cycle'' (LDC) transients.  The classification of each
source is provided in Table~\ref{etime}.  For a some sources (n1-26,
r1-35, r2-61, r2-65, s1-18, s1-27, s1-69) we could not reliably
determine which class they fell into, and no classification is
reported in Table 3.

\subsection{Short Term Variability}\label{stv}

We tested our sources for short-term variability (type I bursts)
through Kolmogorov-Smirnov (KS) tests of the arrival times of the
photons against a constant arrival time distribution.  We applied this
test to the ACIS detection of each source with the highest number of
counts, if that detection contained over 100 counts.  All of these
tests yielded $P$ values greater than 0.75, providing no evidence for
significant short-term variability in any of the transient events.

We tested the sensitivity of the KS test by creating monte carlo
photon arrival time distributions for detections containing 100 counts
distributed over 5 ks, the typical length of our ACIS observations.
We ran such tests for bursts of 10, 100 and 1000~s.  These tests
showed that our observations were more sensitive to longer bursts.
They were sensitive to factor of 2.3 changes in flux that lasted
1000~s, factor of 10 changes in flux that lasted 100~s, and factor of
100 changes in flux that lasted 10~s at the 50\% confidence level.
Galactic X-ray Type~I bursts (see Fig. 3.1 in review by
\citealp{lewin1993}) typically last $\sim$10--100~s with factors of
$\sim$10--20 changes in flux, so that we would be sensitive only to
the brightest and longest bursts.  For example, our data would not be
sensitive to the Type~I bursts in Aql~X-1 \citep{czerny1987}, because
its flux increased by a factor of 10 for less than 100~s.

\subsection{Periodic Outbursts}

We inspected the lightcurves for transients that may be periodic.
Only 3 sources in the sample showed more than 2 distinct outbursts.

Sources r1-11 and r2-16 appear to have undergone $\sim$4 outbursts
from 1999 to 2002. The outbursts appear at regular intervals,
suggesting that these transient sources have outburst periods of
$\sim$0.7 yr and decay time estimates of 40 and 90 days, respectively.
These numbers are comparable to the values of the Galactic neutron
star LMXB transient Aql~X-1, which undergoes a 30--60 day outburst
every $\sim$200 days \citep{simon2002}.  These sources also have some
detections with the hardest spectra in the sample, with photon indexes
as low as 0.8 (r1-11) and 0.5 (r2-16), but they are not expected to be
HMXBs because they are $<$1.3$'$ from the center of M31 where there is
no known young stellar population (e.g. \citealp{stephens2003}).  The
absorption columns for the sources are consistent with $\sim$10$^{21}$
cm$^{-2}$, so that the secondary is not likely heavily extincted.  If
these sources are HMXBs, they have secondaries fainter than
M$_V\sim$-1 (see \S~\ref{galactic}).  In any case, the recurrence and
spectral properties hint that these transients may have neutron star
primaries.

Source r3-8 also showed 4 outbursts, but this source is much softer
than r2-16 and r1-11.  Source r3-8 has equivalent hardness ratios to
several SSSs and therefore is unlikely to be an HMXB.

\subsection{Spectral Properties}\label{spectra}

The results of the photometric and spectroscopic measurements of all
of the transient source detections are given in Table~\ref{spec.dat}.
Histograms of the measured photon indexes ($\Gamma$) and absorption
columns are shown in Figure~\ref{spec.hist}.  Again, a wide range of
spectral properties are seen, from hard spectra like those of s1-1 and
r1-5, which are well-fit by power-law models with $\Gamma \sim 1.5$,
to very soft spectra like those of r3-8 and n1-86, which have hardness
comparable to those of SSSs.

There is also a wide range of absorption column values for these
sources, ranging from 4$\times$10$^{20}$ to 10$^{22}$ cm$^2$.  The
errors on the absorption measurements tend to be larger than those on
the $\Gamma$ measurements, reflecting the effective area of the
telescopes at low energies.  Nevertheless, most of the measured
absorption values for these sources are 0.4$\times 10^{21}$--3$\times
10^{21}$ cm$^2$, typical absorption values for M31 X-ray sources,
suggesting that the vast majority of these measurements are reliable.

Generally, the spectra of the sources appear to vary between
observations, as evidenced by the hardness ratios.  In many cases, the
errors on the spectral parameters are too large to reliably quantify
the spectral changes.  On the other hand, there are some cases where
the changes are large enough to be statistically significant in the
parameters of the spectral fits, such as r2-3, whose photon index
varies between 1.6$\pm$0.2 to 2.7$\pm$0.2, and r3-16 whose index
varies from 1.4$\pm$0.2 to 2.0$\pm$0.2.  Absorption values also can
also be variable in these transient sources, indicating that they are
self-absorbed.  For example, the absorption toward r1-5 varies between
0.7$\pm$0.2 $\times 10^{21}$ cm$^2$ and 2.9$\pm$0.8 $\times 10^{21}$
cm$^2$, and the absorption toward r2-8 varies between 0.3$\pm$0.3
$\times 10^{21}$ cm$^2$ and 2.2$^{+1.1}_{-0.8} \times 10^{21}$ cm$^2$.

\subsection{Neutron Star Atmosphere Fits}\label{nsa}

Some of the softest transients were best-fitted by a neutron star
atmosphere (NSA) spectral model \citep{zavlin1996}.  This model
provides the spectra emitted from the hydrogen atmosphere of a neutron
star.  When fitting this model, we fixed the magnetic field to zero,
the radius to 10 km, and the mass to 1.4 M$_{\odot}$, allowing only
the effective temperature of the atmosphere to be fitted.  Three
sources were well-fitted by this model (n1-86, r3-8, and s1-18);
however, the fluxes supplied by all of the NSA fits indicate (assuming
L$_X = 4\pi R^2 \sigma T_{eff}^4$) that the emitting source has a
radius of $\sim$10$^5$ km if located in M31, making them much larger
than neutron stars and even larger than white dwarf stars.  If they
are neutron stars (10 km radius), they have L$_X\sim
10^{29}$--10$^{30}$ erg s$^{-1}$ and are at a distance of $<$100 pc,
making them Galactic foreground sources.  The local space density of
these sources implied by this result is not consistent with Galactic
observations.  However, if these sources are cataclysmic variables
(CVs), their space density is consistent with the local value.

Source n1-86 is only well-fitted ($\chi^2/\nu < 1.5$) in all
observations with a NSA model; however, a blackbody model provides an
equally good fit for the January 2002 ACIS detection.  Because the NSA
fits required a slightly lower absorption correction and provided the
only acceptable fits to the January 2002 {\it XMM} detection and
February 2002 ACIS detection, we took NSA as the appropriate model for
all detections of this source.  The effective temperature and
luminosity (assuming the source is in M31) that come out of this fit
implies that the radius of the star is $>$10$^5$ km.  Since this
radius is even large for a white dwarf, it is possible that the source
is a Galactic foreground neutron star or CV.  This possibility would
also explain the high luminosity measured for this source
(2$\times$10$^{39}$ erg s$^{-1}$).  If n1-86 is in M31, it had the
highest X-ray luminosity of any transient yet observed in M31.  If it
is a foreground neutron star or CV, its luminosity was $<$10$^{36}$
erg s$^{-1}$.

Source s1-18 only has a single detection useful for spectral fitting.
The small number of counts is well-fitted by both a blackbody model
($\chi^2/\nu = 1.3$) and a NSA model ($\chi^2/\nu = 0.6$).  Because
the NSA fit is better, we took it as the correct model.  In any case,
the absorption-corrected luminosity did not depend significantly on
the choice of model for this source. The effective temperature and
luminosity that come out of this fit implies a large radius of
$>$10$^4$ km, indicating that the source may be in the Galactic
foreground with luminosity $<$10$^{36}$ erg s$^{-1}$.

\begin{figure}
\centerline{\psfig{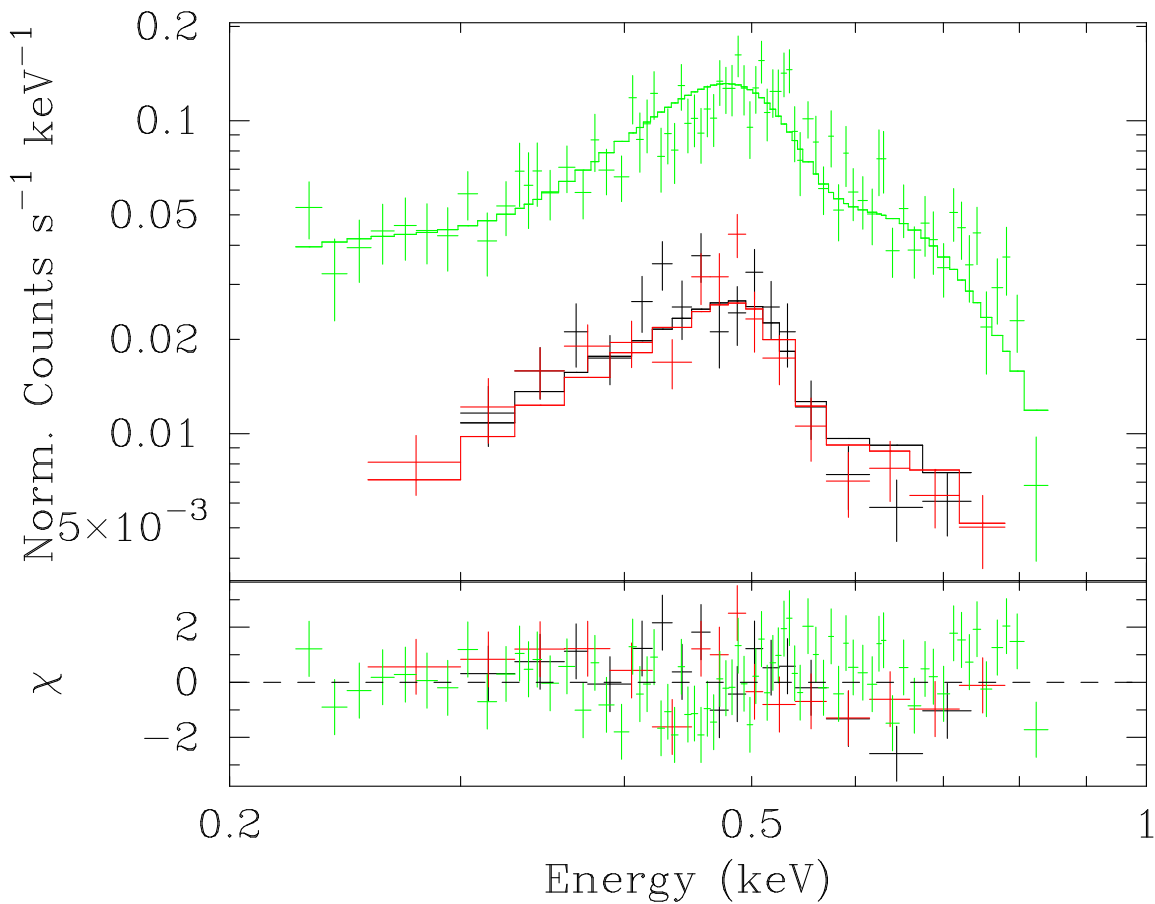}}
\caption{The best-fit single-component model for the 2002 January 06
observation of r3-8. {\it Top:} The green histogram marks the best fit
model prediction for the PN detection and error bars mark the measured
count rates from the PN data. The red histogram and error bars show
the same fit for the MOS 1 data, and the black histogram and error
bars show the same fit for the MOS 2 data. {\it Bottom:} Residuals for
the spectral fits.  The colors correspond to the same instruments as
in {\it top}.  The spectrum was equally well-fit by a multi-component
model (absorbed blackbody plus Gaussian emission line).}
\label{r3-8}
\end{figure}

The most significant improvement for a spectral fit using the NSA
model was for r3-8.  All of the high-count {\it XMM} and {\it Chandra}
detections of this source were well-fitted with the NSA model.  In
addition, the detection with the highest number of counts, from {\it
XMM} in January 2002, could only be fitted with the NSA model.  No
other model, including composite models, fit the source with
$\chi^2/\nu < 2$.  The NSA model fits the spectrum quite well, as
shown in Figure~\ref{r3-8}.  Although quantitatively the fit has
$\chi^2/\nu = 1.3$, it clearly follows the continuum correctly, and no
other single-component model has such a shape to its spectrum, showing
the observed peak at $\sim$0.5 keV and a sharp drop to higher
energies.  The fact that only this model fits the spectrum suggests
that r3-8 is a compact star with an atmosphere effective temperature
of 2$\times$10$^5$ K.  Assuming the source is in M31 provides a
luminosity indicating that the radius of the star is $>$10$^5$ km.
Again, this result suggests the source is in the Galactic foreground
with a luminosity of $<$10$^{36}$ erg s$^{-1}$.

Because these sources appear to be in the Galactic foreground, we
investigated whether such a number of faint foreground transient
sources in our sample is reasonable and what sort of objects could be
the sources of this X-ray emission.  It is possible that such a number
of faint foreground transients has gone undetected until now.  A
surface density of such transient sources, with fluxes of
$\sim$10$^{-13}$--10$^{-12}$ erg cm$^{-2}$ s$^{-1}$, of $\gap$1 per
square degree would not contradict the results of Galactic surveys
such as, for example, the ChaMPlane survey \citep{grindlay2003}, even
though ChaMPlane does not repeat observations many times and therefore
would not be sensitive to such transient sources.

Even though the only single-component spectral model that fits these
data is a NSA model, these sources are not likely to be neutron stars.
With the measured temperatures, neutron stars would have luminosities
of $\sim$10$^{29}$--10$^{30}$ erg s$^{-1}$.  Our survey was only
sensitive to such sources within $\sim$200 pc, yielding a volume
density of $\gap$10$^{-3}$ pc$^{-3}$ for these sources.  Such a volume
density is unreasonable under the assumption that CVs (whose space
density is of order $10^{-5}$ pc$^{-3}$; \citealp{grindlay2005}) are
more common than transient neutron stars.

The sources are too bright to be CVs with spherically symmetric
emission.  The measured temperatures yield spherically symmetric
luminosities of $\gap$10$^{35}$ erg s$^{-1}$ for white dwarf radii.
At these luminosities, the distances are $\gap$10 kpc, putting the
sources far out of the Galactic disk.  These luminosities and
locations are both atypical of CVs.

The most reasonable explanation for these sources is that they are
foreground polars.  In this case, the radii of the primary stars is
$\sim$8000 km \citep{warner1995}, and the fractional emitting surface
area is $\sim$10$^{-3}$ \citep{mauche1999}, making the luminosities of
the sources $\sim$10$^{32}$--10$^{33}$ erg s$^{-1}$
(3$\times$10$^{32}$ erg s$^{-1}$ for n1-86; 7$\times$10$^{32}$ erg
s$^{-1}$ for r3-8; 2$\times$10$^{32}$ erg s$^{-1}$ for s1-18), and
their distances less than a few kpc (300 pc for n1-86; 1 kpc for r3-8;
2 kpc for s1-18).  These temperatures, luminosities and distances
would be reasonable for polars (see \citealp{warner1995} and
\citealp{cropper1990} reviews).  Furthermore, with these distances and
luminosities, the space density of the sources would be
$\gap$4$\times$10$^{-7}$ pc$^{-3}$, which is similar to several
published values for the space density of polars
\citep{cropper1990,warner1995}.

It is unusual that the spectra of these sources are well-fit by an NSA
model if they are not neutron stars, but we note that the NSA model
was simply the only single-component model with which we could fit the
data.  It is possible that multi-component models with more free
parameters, such as those used to fit the spectra of polars, could
also fit these data.  There are similarities between the spectra of
our NSA sources and those of known polars.  For example, these sources
are all very soft, and some polars are known to have strong soft
emission \citep{ramsay2005fig5}.  Furthermore, although typical XMM
spectra of polars in the bright phase have spectra that are flat or
decreasing from 0.2--1 keV (see figures in
\citealp{ramsay2001,ramsay2004}), spectra similar to that observed for
r3-8 do exist (e.g., RX J1914+24; \citealp{ramsay2005rx1914}).

Many polars have spectra that are difficult to fit and require
modeling with multi-component blackbody spectra.  Our XMM spectrum of
r3-8 provided enough counts to attempt a fit with such multi-component
models.  An absorbed ($N_H = 2.3 \times 10^{21}$ cm$^{-2}$) blackbody
spectrum with kT = 71 eV plus a broad Gaussian emission line at 0.52
keV, similar to the best-fit model of the polar RX~J1914+24
\citep{ramsay2005rx1914}, provides a good fit with $\chi^2/\nu =
124.5/98$.  We therefore suggest, based on the space density and
spectra, that these sources are Galactic foreground polars and that
the spectra of polars can mimic those of NSAs.  More observations of
these very soft transient sources are needed to confirm this
possibility.

\subsection{Hardness-Luminosity and Color-Color Diagrams}

\begin{figure}
\centerline{\psfig{figure=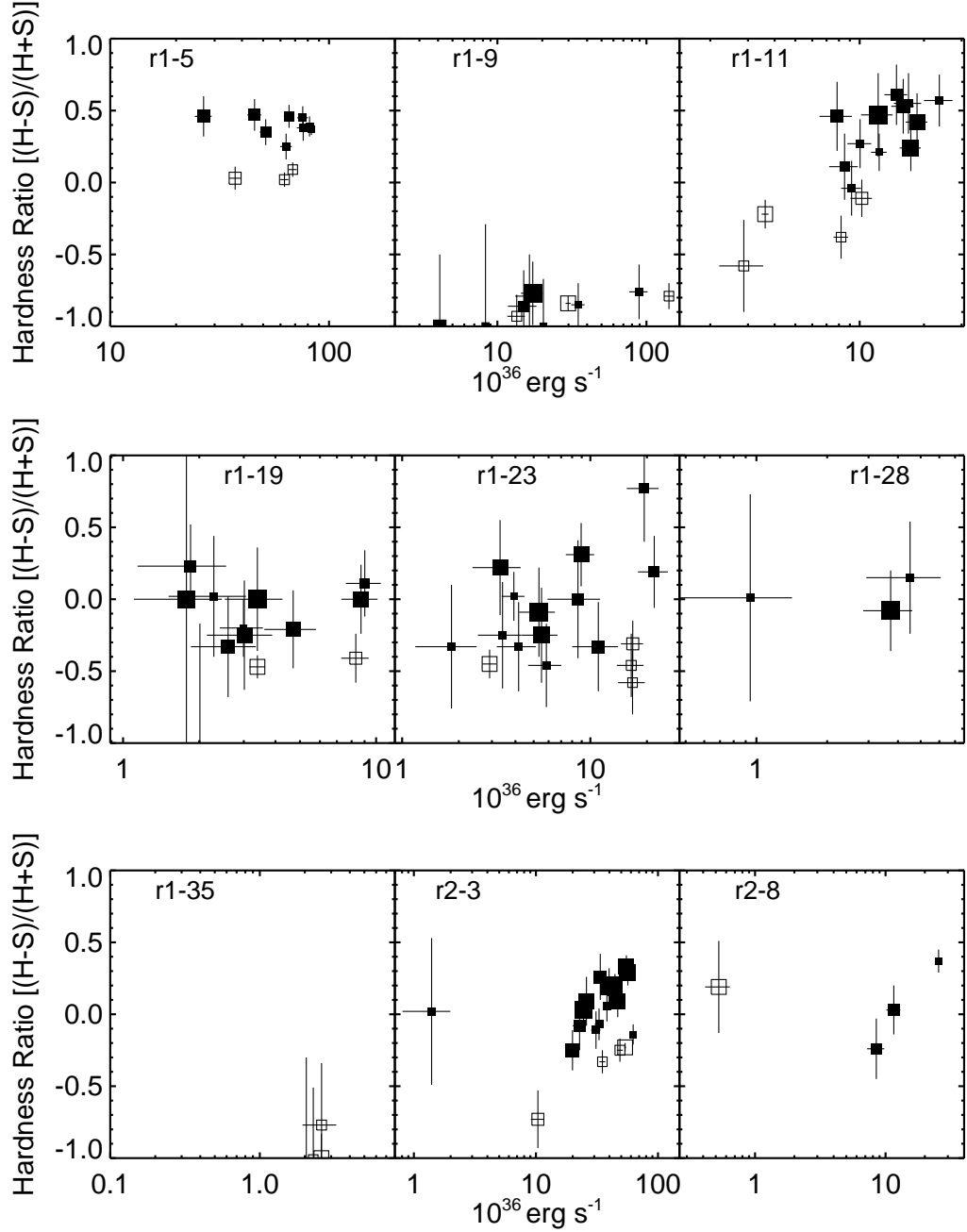,height=7in,angle=0}}
\caption{The hardness-luminosity diagrams for nine transient sources
detected 3 or more times with the ACIS detectors.  Filled squares are
ACIS-I data points and hollow squares are ACIS-S data points.  The
luminosities are absorption corrected 0.1--10 keV luminosities
measured using the most contemporaneous spectral fit available, and
the hardness ratios [(H-S)/(H+S)] are from background-subtracted
source counts in 2 bands: S (0.3--1.0 keV) and H (2.0--7.0 keV).  The
sizes of the data points reflect the times of the observations.
Larger data points represent later observations.}
\label{hi9}
\end{figure}

\begin{figure}
\centerline{\psfig{figure=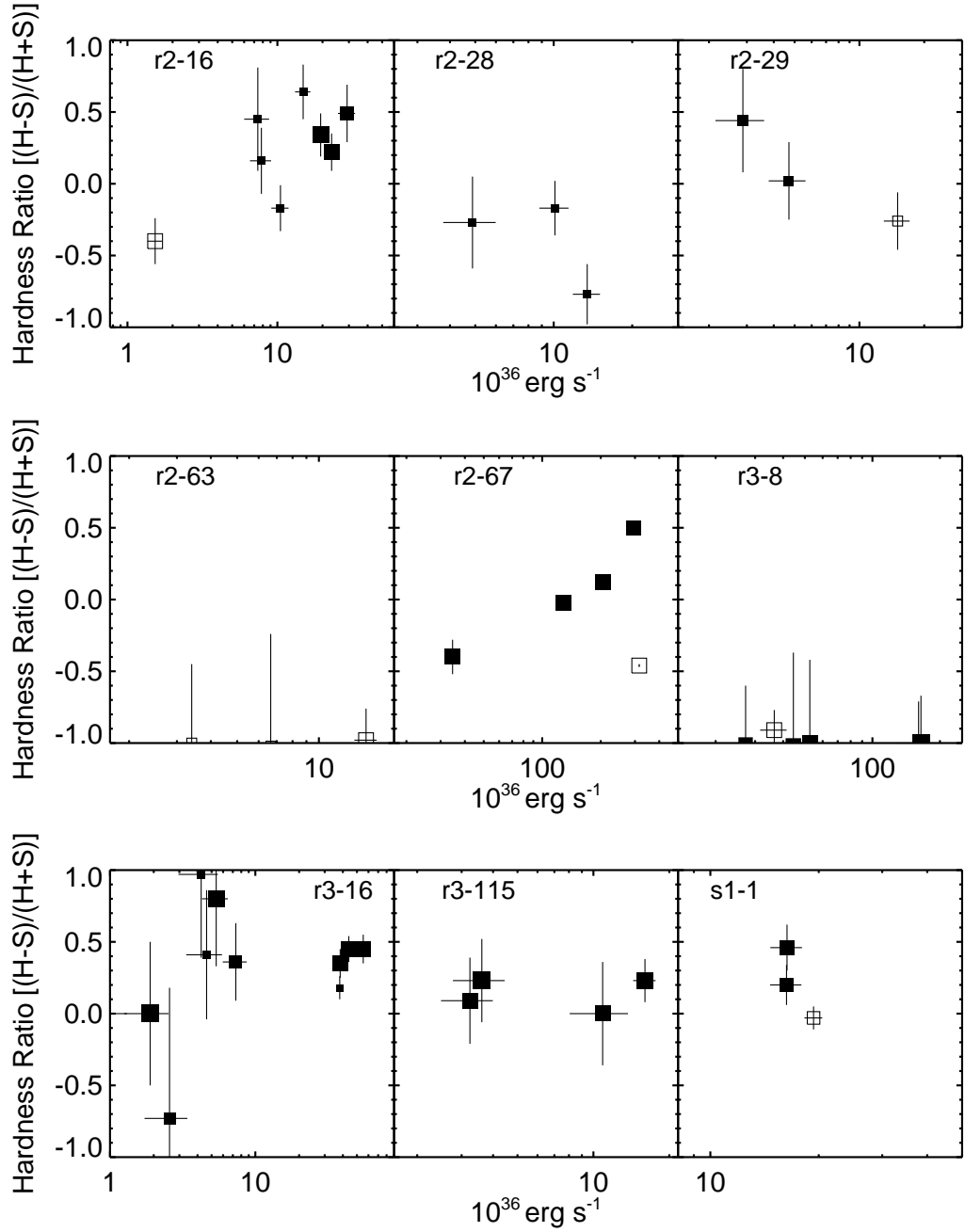,height=7in,angle=0}}
\caption{The hardness-luminosity diagrams for nine more transient
sources detected 3 or more times with the ACIS detectors.  Data
representation is identical to that in Figure~\ref{hi9}.}
\label{hi18}
\end{figure}

To monitor the spectral properties of the transient events, we
produced hardness-luminosity (hardness-intensity proxy) diagrams from
the hardness ratio HR2 and absorption-corrected 0.3--7 keV
luminosities from the ACIS-I and ACIS-S detections of the transient
events.  These diagrams were only generated for events with at least 3
total ACIS detections.  This requirement limited the number of
diagrams to 18, all of which are shown in Figures~\ref{hi9} and
\ref{hi18}. We have plotted the ACIS-S and ACIS-I ratios with
different symbols because the ratios are calculated from straight
counts.  Because ACIS-S is more sensitive to soft photons than ACIS-I,
the ACIS-S ratios will be systematically softer than those from ACIS-I
for the same source.

These diagrams show a wide range of behavior.  For example, r1-9,
r1-35, r2-63, and r3-8 exhibit very soft ratios at all luminosities,
providing only upper-limits to their hardness ratio in many cases.
This behavior clearly sets them apart as SSSs. In addition, the
diagrams of some sources, such as r1-5, r1-23, r2-3, r2-8, r2-16,
r2-67, and r3-16, show interesting patterns.

The diagram for r1-5 reveals a range of luminosities with a hard
spectrum with a hint of a transition from the bright, hard state to a
soft state at similar luminosity. Note that the ACIS-S ratios are
softer because of the higher sensitivity to soft photons.  Source
r1-23 appears to vary between a low-soft and high-hard state.  Source
r2-3 reveals a range of hardness ratios that is consistent with a
transition from a low-soft state to a high-hard state, with a single
data point suggesting an extension to a low-hard state.  The bimodal
HR diagram looks similar to that for Galactic Z-sources, suggesting we
are seeing the Normal Branch and Flaring Branch but missing the
Horizontal Branch (see Figure 3.1 of \citealp{kuulkers1995}). The
diagram for r2-16 shows an almost circular pattern, going from a
low-soft state, to a low-hard state, to a high-hard state, to a
high-soft state.  This pattern is also seen in some Galactic
Z-sources, such as Cyg~X-2, (see Figure 4.3 of
\citealp{kuulkers1995}).

At the same time, there are several similarities with Galactic black
hole binaries.  For example, r2-3 and r2-8 show transitions from a
high-hard state to a low-soft state.  Source r2-67 appears to show the
same kind of transition, although we note that some of this effect
could be due to pileup during the high luminosity observations
hardening the spectrum.  The pattern is reminiscent of the most
luminous portions of the diagram for XTE~J1859+226 and GRO~1655-40,
which show their softest colors at intermediate fluxes
\citep{done2003}. This comparison makes sense because we only detect
the M31 transients when they are bright.  Source r3-16 has very large
errors in hardness ratio at low luminosities, but it appears to
undergo a similar transition near its peak luminosity.  In addition,
the hardness luminosity diagram for r1-19 shows a slightly softer
color at 2--4$\times$10$^{36}$ erg s$^{-1}$ than at brighter and
fainter luminosities, and that of r2-16 shows a strong dip in hardness
at $\sim$10$^{37}$ erg s$^{-1}$.  This behavior appears comparable to
that of XTE J1550-564 from $\sim$0.001--0.005 L$_{\rm Edd}$ (see Fig.2
of \citealp{done2003}).  Several sources appear soft when they are
brightest, such as r2-28 and r2-29, also similar to many Galactic
black holes.

\begin{figure}
\centerline{\psfig{figure=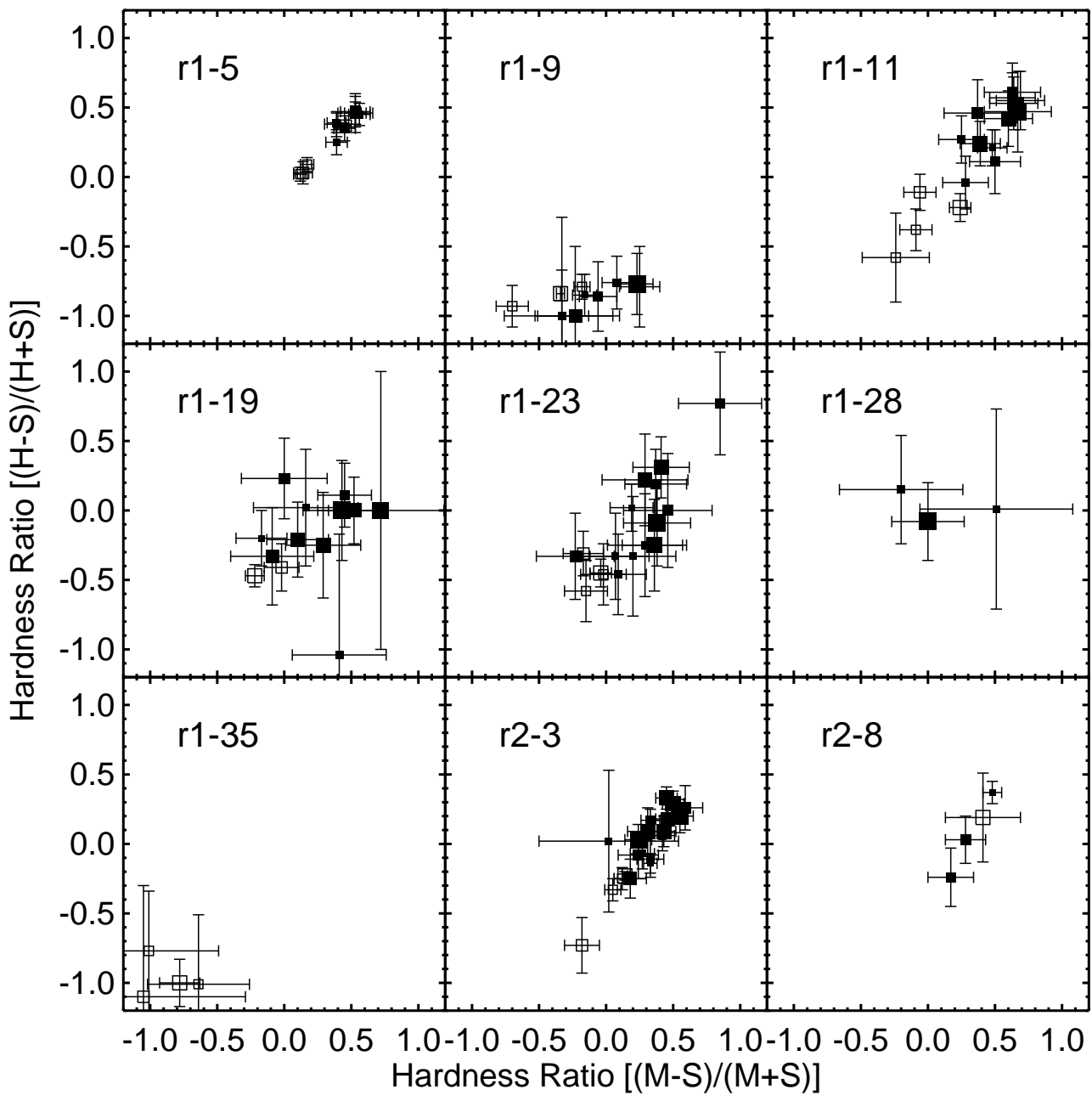,height=7in,angle=0}}
\caption{The color-color diagrams for nine transient sources detected
3 or more times with the ACIS detectors.  Filled squares are ACIS-I
data points and hollow squares are ACIS-S data points. The hardness
ratios are from background-subtracted source counts in 2 bands: S
(0.3--1.0 keV) and H (2.0--7.0 keV).  The sizes of the data points
reflect the times of the observations.  Larger data points represent
later observations.}
\label{cc9}
\end{figure}

\begin{figure}
\centerline{\psfig{figure=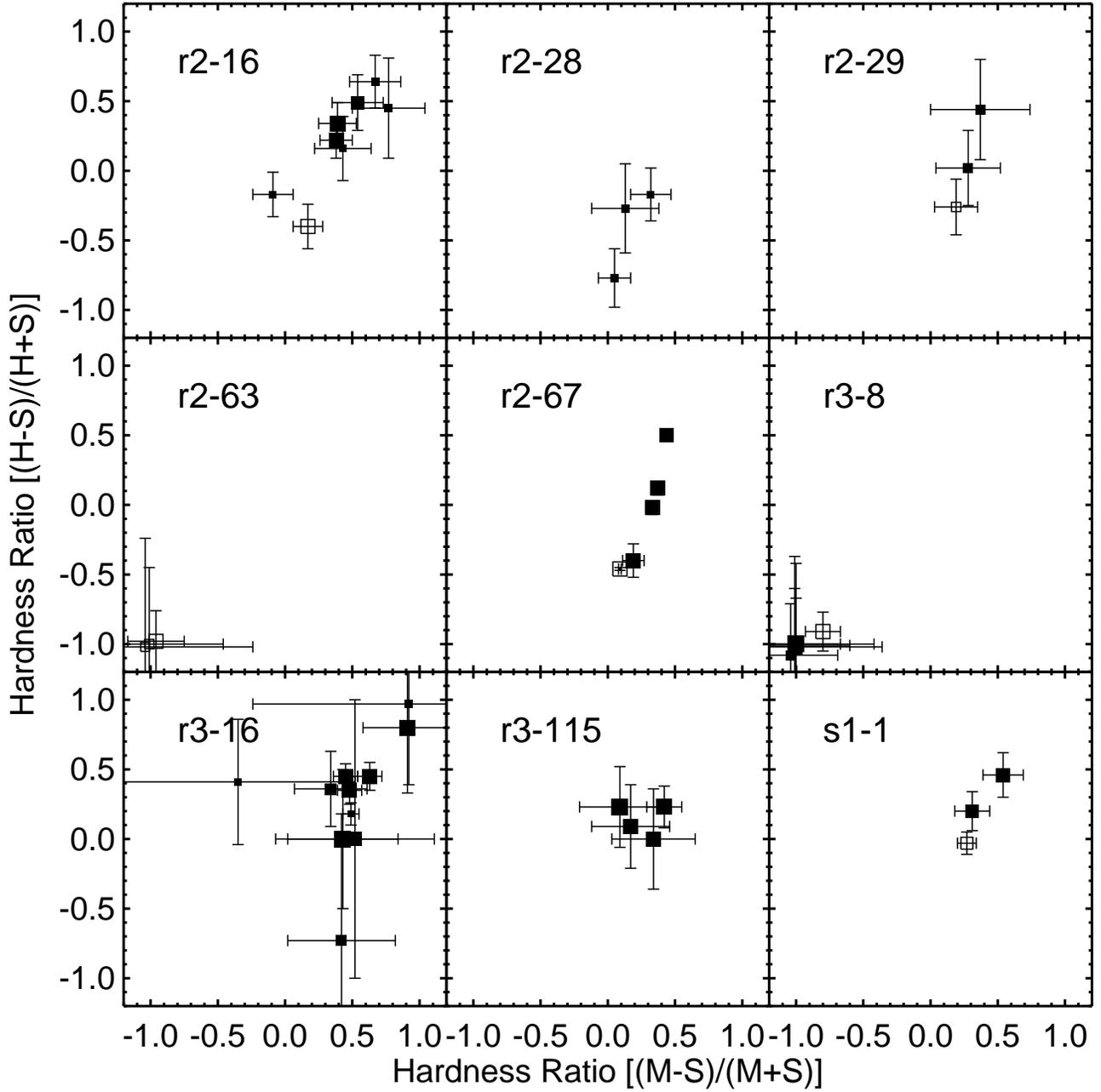,height=7in,angle=0}}
\caption{The color-color diagrams for nine transient sources detected
3 or more times with the ACIS detectors.  Data representation is
identical to that in Figure~\ref{cc9}.}
\label{cc18}
\end{figure}

We also created color-color diagrams of these sources, shown in
Figures~\ref{cc9} and ~\ref{cc18}.  When these types of diagrams are
made for bright Galactic LMXBs, they tend to fall into two categories:
``Z'' sources and atoll sources \citep{hasinger1989}.  The color-color
diagrams for ``Z'' sources form a the shape of a ``Z'' as they evolve
from hard to soft and back to hard; the others form ``atolls'', where
they jump more quickly from hard to soft and back.  The color-color
diagrams of several of sources that are not supersoft (r1-9, r1-35,
r2-63, r3-8 are SSSs) resemble ``Z'' sources, such as r1-5, r1-11,
r1-23, r2-3, r2-8, and r2-16.  For example, the color-color and
hardness-intensity diagrams for r2-16 resemble those of Cyg X-2 in the
Normal Branch and Flaring Branch \citep{kuulkers1995}, but this linear
track from soft to hard colors is also seen in black hole binaries
\citep{done2003}, making it difficult to distinguish between the two
possibilities.  Source r3-16 has a distinctive color-color diagram
that appears to follow a linear track at hard colors but then breaks
up at soft colors, similar to tracks seen in the color-color diagrams
of some Galactic black holes, such as XTE~J1550-564.

\section{Discussion}\label{discussion}

\subsection {Galactic Comparisons}\label{galactic}

Transient X-ray sources have been studied in detail in the
Galaxy. Therefore comparisons between the general spectral and
photometric properties of the Galactic and M31 populations may aid in
drawing conclusions about the types of X-ray binaries causing the
outbursts observed in M31.  In addition, any differences between the
M31 outbursts and those observed in the Galaxy will raise new
questions about the M31 X-ray binary population.  It has already been
suggested that the LMXB population in the M31 bulge contains roughly
the same BH/NS ratio as that seen in the Galaxy \citep{williams2004}.
Now we would like to discover if the M31 XRTs themselves share similar
physical properties to their Galactic siblings.  Bright X-ray
transient events originate from several different types of objects.
These include supersoft sources (SSSs), high-mass X-ray binaries
(HMXBs), and low-mass X-ray binaries (LMXBs).

The physical nature of SSSs is not well-understood, but they are
straight-forward to identify, as their outbursts are relatively faint
($\lap$10$^{37}$ erg s$^{-1}$), and their X-ray spectral properties
are supersoft, with $\Gamma \gap 3.5$ or blackbody temperatures of
$\sim$100 eV.  SSSs are often seen as recurrent X-ray transients
(e.g. \citealp{white1995}).  The most widely accepted model for SSSs
consists of a white dwarf accreting from a companion at a high enough
rate ($\gap$10$^{-7} M_{\odot} yr^{-1}$) to sustain stable nuclear
burning \citep{vandenheuvel1992,rappaport1994}.  However, the
definition of the class is broad enough that NS binaries and isolated
stellar core models have been considered
\citep{greiner1991,distefano2001}.

There are several sources in our catalog that could be SSSs.  Sources
n1-26, r1-9, r1-35, r2-60, r2-61, r2-62, r2-63, r2-65, r2-66, r3-115,
s1-18, s1-27, s1-69, and s2-62 are already classified as QSSs and SSSs
\citep{distefano2001,distefano2004}.  In addition, r3-128 has a
spectrum that is not well fit by any model, but the best fit is a very
low temperature (0.04 keV) thermal model, suggesting it may be
supersoft.  Finally, r3-8 and n1-86 have hardness ratios consistent
with typical SSSs and their spectra are equally difficult to fit, only
providing good $\chi^2$ values for a low-temperature NSA model.

HMXB transient X-ray events typically occur near star forming regions,
as they contain high-mass donor stars.  These donor stars are
typically Be stars \citep{tanaka1996}.  These stars have spectral
types between O8.5 and B2 \citep{negueruela1998} and optical
magnitudes of -4.8$\lap$M$_V\lap$-2.4 \citep{cox2000}.  The X-ray
spectra of these outbursts tend to be harder than the other types of
XRTs, with photon indexes near 1.  In addition, these systems tend to
have periodic repeating outbursts thought to occur during close
encounters between the high-mass donor star and the neutron star
primary \citep{tanaka1996}.  Two of our sources have outbursts that
appear to be periodic, r1-11 and r2-16.  These two sources are the
most likely HMXBs in the bulge.  However, inspection of the 2004
January 23 {\it HST} ACS observation described in
\citet{williams2005bh4}, reveals no bright stars ($B\lap23.5$;
M$_V\lap$-1) within an arcsec of the X-ray positions of r1-11 or
r2-16.  Only r3-16 and r3-115 have bright, persistent optical
counterparts \citep{williams2004,distefano2004}.  Finally the XRT
spectra are nearly all soft, so that very few of the sources in our
catalog are likely to be HMXBs.  Outside the bulge, n1-87 and s1-1
have spectral properties that are consistent with a HMXBs; however, we
note that several sources outside the bulge do not have spectral data
available.

Neutron star LMXBs that undergo bright (10$^{38}$ erg s$^{-1}$), long
(decay time $>$10 day) X-ray outbursts are rare, possibly due to the
presence of a hard stellar surface
(e.g. \citealp{king1997,kingandkolb1997}).  Only a handful of such
transient X-ray sources are known \citep{mcclintock2003}.  Examples of
such systems are Aql~X-1 \citep{koyama1981} and Cen X-4
\citep{matsuoka1980}, which were shown to contain neutron stars by the
presence of type I X-ray bursts in their lightcurves.  In addition to
their type I bursts, these XRTs have longer outbursts that recur on
timescales of years, mimicking the bursts of BHBs because they have
X-ray luminosities of up to $\sim$2$\times$10$^{38}$ erg s$^{-1}$ and
decay times of about 1 month.  Although our data are not sufficient to
reliably determine how many sources in our catalog may be neutron star
LMXB systems, the small number of known neutron star XRNe in our
Galaxy suggests that there are at most only a handful.  Furthermore,
our short-term variability tests were sensitive to the bright-end of
type I bursts (factor of $\sim$10 flux increases with $\sim$100~s
duration; see \S~\ref{stv}), and no short-term variability was
detected in our sample.

Some neutron star LMXBs are in ultra-compact X-ray binaries (UCXBs),
which tend to form in areas of high stellar density, such as the cores
of globular clusters or near the centers of galaxies.  A population of
these sources has been observed in the Galactic center
(e.g. \citealp{heise1999,king2000}). These sources are believed to
have neutron star primaries.  Their X-ray spectra are typical of LMXBs
containing neutron star primaries, power-law with $\Gamma \sim
1.7$. Their short orbital periods ($\lap$2 hr) show that they are
close binary systems.  The small orbital separation restricts the size
of the accretion disk, resulting in short outburst decay times and
faint outbursts.  These systems have decay times of only $<$1 week,
and peak luminosities of $\lap$10$^{37}$ erg s$^{-1}$.  Their rapid
variations occur on the time scale of a few days, as in 4U 1820-30
\citep{simon2003}.  In addition, they can undergo ``superbursts'' with
luminosities of $>$10$^{38}$ erg s$^{-1}$, as in 4U 1820-30, but the
decay time is $<$1 day \citep{strohmayer2002}. Four sources (r2-61,
r3-43, r3-128, and s1-69) were detected only once, allowing the
possibility of such a short decay time.  In addition, there are five
sources (n1-59, n1-85, r2-65, r2-69, and r3-8) that were detected
once, then were not detected in the following observation but
reappeared two or three observations later for a single observation.
Three of these 9 sources (r2-61, r2-65, and s1-69) have previously
been classified SSSs by others \citep{distefano2004}, and one source
(r3-8) may be a foreground CV (see \S~\ref{nsa}). Therefore, five
sources have the potential to be UCXBs.  Of these five, the most
likely are r2-69, r3-43, and r3-128 because they are located in the
high stellar density M31 bulge.

Succinctly, many soft Galactic XRT events have BH primaries.  These
XRTs share many properties with the few neutron star LMXB transient
events described above, including long recurrence timescales, outburst
decay timescales of $\sim$1 month, and X-ray luminosities.  However,
there are some differences, including softer X-ray spectra, with
$\Gamma > 1.7$ and/or temperatures of $\lap$1 keV.  Although their
outbursts can be as faint as 10$^{36}$ erg s$^{-1}$, some of these
systems have outbursts $>2\times10^{38}$ erg s$^{-1}$, ruling out
neutron star progenitors.  Radial velocity studies of Galactic XRT
sources with these properties suggests $\sim$90\% of them are BHBs
\citep{mcclintock2003}.  It is therefore likely that most of the
sources in our catalog, excluding the SSSs, are BHBs.

\section{Conclusions}\label{conclusions}

We have put together all of the {\it Chandra} and {\it XMM-Newton} data sets
pertaining to 45 transient X-ray sources in M31 detected from October
1999 to August 2002.  These data have provided a catalog of positions,
spectral properties, and lightcurves for transient X-ray sources in
M31.

The locations of these transient sources indicate that they most often
form near the center of the galaxy, suggesting that most of them are
old LMXBs.  The distribution of peak luminosities of these sources
range from $\sim$10$^{36}$ to $\sim$10$^{39}$ \ergs , similar to
Galactic XRTs.  The estimated $e$-folding decay times are similar to
those seen in Galactic XRNe, also suggesting that many of these
sources are BH-containing LMXBs.

The transient source lightcurves and positions indicate 5 potential
UCXBs, 3 of which are located in the M31 bulge.  Two other sources
have possible periodic outbursts and hard spectra consistent with
HMXBs, but they have no optical counterparts brighter than M$_V
\approx -1$.

The lightcurves also constrain the transient rate in M31.  If we take
only those sources with the best lightcurves (Hi/Lo $>$15, measured
decay time, and one at least one clear outburst), there were 26
transient sources seen in the 35 months covered by our sample.
Removing the total of $\sim$9 months during which M31 was not
observable during this time period yields a transient rate in M31 of
$\sim$1 per month, consistent with the previous result in
\citet{williams2004}.  The rate of X-ray novae in the Galaxy
($\sim$2.6 per year; \citealp{chen1997}), hinting that the rate in M31
could be higher.  However, since the Galactic rate is likely
incomplete because sampling the entire Galaxy is much more difficult
than sampling M31, the similarity of the two rates is noteworthy.

Furthermore, these data are some of the best available for studying
the duty cycles of a large sample of transient sources.  We have used
the lightcurves to constrain the duty cycles of the transients,
finding about half have duty cycles $<$0.1 and all but 3 sources have
duty cycles $<$0.2.  This range is similar to that seen in the Galaxy
\citep{chen1997}, where the transients have duty cycles ranging from
0.0005--0.3 and most duty cycles are $<$0.1.

The spectra of the sources provided more details about their physical
properties.  Some source spectra are well-fitted by simple absorbed
power-law models with photon indexes ranging from 1--4 and absorption
values from 0.1--10 $\times$ 10$^{21}$ cm$^{-2}$.  Others are only
well-fitted by multi-component spectral models, including power-law
plus blackbody or disk blackbody models with temperatures 0.03--1.2
keV.  Still other source spectra are too soft to provide an acceptable
fit with any of these model combinations.  Three soft sources are
well-fit by a neutron-star atmosphere model, but the best-fit
effective temperatures indicate that these sources must be in the
Galactic foreground.  The surface density of these sources ($\gap$1
deg$^{-2}$) is consistent with that of magnetic CVs (polars) but
higher than expected for neutron stars, suggesting that these sources
are foreground polars. Their spectra are also well-fit by
multi-component models similar to those of some polars, showing that
polar spectra and neutron star atmosphere spectra can be similar.

Clearly the discovery of these XRTs marks only the beginning of their
study.  These sources provide the largest extra-galactic sample of
XRTs available in one galaxy and will open the door to future
comparison studies of outbursting X-ray binary populations in
galaxies.

We thank Robin Barnard for suggesting the NSA model for fitting the
spectra of some of the soft sources.  Support for this work was
provided by NASA through grant GO-9087 from the Space Telescope
Science Institute and through grant GO-3103X from the {\it Chandra}
X-Ray Center.


\small
\footnotesize
\clearpage
\footnotesize
\begin{deluxetable}{cccc}
\tablewidth{6.5in}
\tablecaption{Transient Catalog}
\tableheadfrac{0.01}
\tablehead{
\colhead{\bf{Source}} &
\colhead{\bf{RA (J2000)}} &
\colhead{\bf{DEC (J2000)}} &
\colhead{\bf{Detection References}}
}
\startdata
n1-26 & 0 46 39.0 & 41 39 07.5 & \citealp{distefano2004}\\
n1-59 & 0 45 45.76 & 41 50 30.1 & \citealp{williams2004}\\
n1-85 & 0 45 28.05 & 41 54 10.9 & \citealp{williams2004}\\
n1-86 & 0 44 14.0 & 41 22 04 & \citealp{iauc7798}\\
n1-87 & 0 44 15.9 & 41 30 58 & \citealp{iauc7798}\\
n1-88 & 0 44 28.4 & 41 56 28.1 & \citealp{atel79}\\
n1-89 & 0 44 38.1 & 41 45 11 & \citealp{iauc7798}\\
r1-5 & 0 42 42.16 & 41 16 08.42 & \citealp{trudolyubov2001}/\citealp{kong2002}\\
\enddata
\label{trans.tab}
\end{deluxetable}

%
\large
\clearpage

\begin{landscape}
\clearpage

\begin{deluxetable}{ccccccccccccccc}
\tablewidth{8.5in}
\tablecaption{Spectral Fits to Transient Detections}
\tableheadfrac{0.01}
\tablehead{
\colhead{\tiny{\bf{Source}}} &
\colhead{\tiny{\bf{Date}}} &
\colhead{\tiny{\bf{Inst}}} &
\colhead{\tiny{\bf{OBSID}}} &
\colhead{\tiny{\bf{Counts\tablenotemark{a}}}} &
\colhead{\tiny{\bf{Model\tablenotemark{b}}}} &
\colhead{\tiny{\bf{${\rm N_H}$\tablenotemark{c}}}} &
\colhead{\tiny{\bf{$\Gamma$\tablenotemark{d}}}} &
\colhead{\tiny{\bf{${\rm kT or log(T)}$\tablenotemark{e}}}} &
\colhead{\tiny{\bf{$\chi^2/\nu$}}} &
\colhead{\tiny{\bf{Q\tablenotemark{f}}}} &
\colhead{\tiny{\bf{HR-1\tablenotemark{g}}}} &
\colhead{\tiny{\bf{HR-2\tablenotemark{h}}}} &
\colhead{\tiny{\bf{$\rm L_X$\tablenotemark{i}}}} 
}
\tablenotetext{a}{\tiny The number of net counts in the detection.  XMM
detection counts are given for 0.3--7 keV in PN/MOS1/MOS2 format.}

\tablenotetext{b}{\tiny The model used to fit the spectrum: pl (power-law),
bb (blackbody), (pb) power-law plus blackbody, (dbb) disk blackbody,
(nsa) neutron star atmosphere.}

\tablenotetext{c}{\tiny The absorption value in units of 10$^{21}$
cm$^{-2}$}

\tablenotetext{d}{\tiny Slope of the best-fitting power law component.}

\tablenotetext{e}{\tiny For the bb or pb models, this number is kT in keV.
For the dbb model this value is kT in keV of the inner edge of the
accretion disk.  For the nsa model this value is $log(T_{eff})$ for
the atmosphere of the neutron star.}

\tablenotetext{f}{\tiny The probability that this fit is representative of
the true spectrum, determined from $\chi^2/dof$.}

\tablenotetext{g}{\tiny Hardness ratio calculated by taking the ratio of M-S/M+S, where S is the number of counts from 0.3--1 keV and M is the number of counts from 1--2 keV.}

\tablenotetext{h}{\tiny Hardness ratio calculated by taking the ratio of
H-S/H+S, where S is the number of counts from 0.3--1 keV and H is the
number of counts from 2--7 keV.}

\tablenotetext{i}{\tiny The absorption-corrected luminosity of the source in
units of 10$^{36}$ \ergs (0.3--7 keV), assuming a distance of 780
kpc.}
\normalsize
\startdata
{\tiny n1-26 } & {\tiny  08-Mar-2001 } & {\tiny  ACIS-S } & {\tiny 2053 } & {\tiny  7 } & \nodata & \nodata & \nodata & \nodata & \nodata & \nodata & {\tiny 0.33$\pm$0.43} & {\tiny -0.35$\pm$0.62} & {\tiny 0.5$\pm$0.2}\\
{\tiny n1-26 } & {\tiny  03-Jul-2001 } & {\tiny  ACIS-S } & {\tiny 2054 } & {\tiny  21 } & \nodata & \nodata & \nodata & \nodata & \nodata & \nodata & {\tiny -0.35$\pm$0.32} & {\tiny -1.12$\pm$0.73} & {\tiny 1.1$\pm$0.4}\\
{\tiny n1-59 } & {\tiny  26-Jan-2002 } & {\tiny  XMM } & {\tiny    0109270301 } & {\tiny 90/\nodata/\nodata} & {\tiny pl} & {\tiny 0.1$^{+1.1}_{-0.1}$} & {\tiny 1.8$^{+1.2}_{-0.4}$} & \nodata & {\tiny 16.61/21} & {\tiny  0.734 } & {\tiny -0.22$\pm$0.23} & {\tiny -0.63$\pm$0.41} & {\tiny 1.1$^{+0.9}_{-0.2}$}\\
{\tiny n1-86 } & {\tiny  05-Jan-2002 } & {\tiny  XMM } & {\tiny    0109270701 } & {\tiny \nodata/318/208} & {\tiny nsa} & {\tiny 2.9$^{+0.7}_{-0.4}$} & \nodata & {\tiny 5.140$^{+0.030}_{-0.050}$} & {\tiny 28.94/29} & {\tiny  0.468 } & {\tiny -1.01$\pm$0.09} & {\tiny -0.98$\pm$0.12} & {\tiny 677.0$\pm$445.0}\\
{\tiny n1-86} & {\tiny 08-Jan-2002} &  {\tiny  ACIS-I }  & {\tiny 2897} & {\tiny  129 } & {\tiny nsa} & {\tiny 2.8$\pm$0.5} & {\tiny \nodata} & {\tiny 5.141$\pm$0.026} & {\tiny 9.28/10} & {\tiny 0.506} & {\tiny -1.03$\pm$0.16} & {\tiny -1.15$\pm$0.18} & {\tiny 623.4$\pm$65.5}\\
{\tiny n1-86} & {\tiny 06-Feb-2002} &  {\tiny  ACIS-I }  & {\tiny 2896} & {\tiny  84 } & {\tiny nsa} & {\tiny 3.9$\pm$1.8} & {\tiny \nodata} & {\tiny 5.219$\pm$0.058} & {\tiny 8.65/6} & {\tiny 0.194} & {\tiny -0.97$\pm$0.19} & {\tiny -0.75$\pm$0.16} & {\tiny 2018.4$\pm$246.6}\\
{\tiny n1-87 } & {\tiny  05-Jan-2002 } & {\tiny  XMM } & {\tiny  0109270701 } & {\tiny 418/129/\nodata} & {\tiny pl} & {\tiny 10.5$^{+3.1}_{-2.1}$} & {\tiny 1.2$^{+0.2}_{-0.2}$} & \nodata & {\tiny 40.97/36} & {\tiny  0.262 } & {\tiny 0.84$\pm$0.14} & {\tiny 0.94$\pm$0.08} & {\tiny 10.5$\pm$2.3}\\
\enddata
\label{spec.dat}
\end{deluxetable}

\clearpage

\begin{deluxetable}{cccccccccccc}
\tablewidth{8.5in}
\tablecaption{Lightcurve Properties}
\tableheadfrac{0.01}
\tablehead{
\colhead{{Source}} &
\colhead{{Peak Date\tablenotemark{a}}} &
\colhead{{$L_{X}\tablenotemark{b}$}} &
\colhead{{$UL\tablenotemark{c}$}} &
\colhead{{$Hi/Lo\tablenotemark{d}$}} &
\colhead{{Decay Start\tablenotemark{e}}} &
\colhead{{$N_{LC}\tablenotemark{f}$}} &
\colhead{{$\tau_d$\tablenotemark{g}}} &
\colhead{{$\chi^2_{\nu}$\tablenotemark{h}}} &
\colhead{{N\tablenotemark{i}}} &
\colhead{{DC\tablenotemark{j}}} &
\colhead{Type}
}
\tablenotetext{a}{\footnotesize The date of the highest measured source luminosity.}
\tablenotetext{b}{\footnotesize The 0.3--7 keV luminosity of the X-ray source at the date given in column 2 in units of 10$^{36}$ erg s$^{-1}$. }
\tablenotetext{c}{\footnotesize The lowest 0.3--7 keV luminosity upper-limit in the lightcurve of the X-ray source in units of 10$^{36}$ erg s$^{-1}$.}
\tablenotetext{d}{\footnotesize The ratio of the highest luminosity of the X-ray source observed to the lowest luminosity upper-limit in the lightcurve.  The highest luminosity observed can be different from the value in column 3 if the highest luminosity observed did not have a well-sampled decay curve.}
\tablenotetext{e}{\footnotesize The date chosen as the beginning of the source decay curve.  Sometimes later than the peak date if the source did not steadily decay after the peak detection.}
\tablenotetext{f}{\footnotesize The number of data points in the decay curve used to estimate the $e$-folding decay time of the source.}  
\tablenotetext{g}{\footnotesize The estimate of the $e$-folding decay time of the source in days.}   
\tablenotetext{h}{\footnotesize The $\chi^2_{\nu}$ value of the fit to an exponential decay adopting $3 \sigma$ errors for detections and $1 \sigma$ errors for upper-limits.}
\tablenotetext{i}{\footnotesize The number of distinct outbursts observed.}
\tablenotetext{j}{\footnotesize The duty cycle of the source is estimated for sources with Hi/Lo $>$ 15, measured decay times, and at least one clear outburst.  Upper-limits for the duty cycle are give for sources that only showed one outburst.}   
\normalsize
\startdata
n1-26 & 2001-07-03 & 1.1$\pm$0.4 & 0.66 &  2 & 2001-07-03 & \nodata & \nodata &  \nodata  & \nodata & \nodata &   QSS\\
n1-59 & 2000-10-12 & 18.8$\pm$7.9 & 0.84 & 22 & 2000-10-12 & 3 & 5 &  1.0  & 2 & 0.03 &   LDC\\
n1-85 & 2001-10-31 & 62.8$\pm$12.6 & 0.16 & 397 & 2001-10-31 & 5 & 8 &  1.8  & 1 & $<$0.02 &   LDC\\
n1-86 & 2002-02-06 & 2018.4$\pm$246.6 & 25.54 & 79 & 2002-02-06 & \nodata & \nodata &  \nodata  & 1 & \nodata &   LDC\\
n1-87 & 2001-10-31 & 29.3$\pm$27.6 & 10.48 &  3 & 2001-10-31 & 4 & 50 &  1.5  & 1 & \nodata &   LDC\\
n1-88 & 2001-11-19 & 14.7$\pm$6.6 & 0.32 & 45 & 2001-11-19 & 4 & 14 &  3.0  & 1 & $<$0.04 &   LDC\\
n1-89 & 2000-05-26 & 9.6$\pm$4.7 & 3.30 &  3 & 2000-05-26 & 5 & 110 &  0.3  & \nodata & \nodata &   LDC\\
r1-5 & 2000-01-19 & 200.5$\pm$24.0 & 0.70 & 285 & 2000-01-19 & 21 & 200 & 21.7  & 1 & $<$0.47 &   LD\\
\enddata
\label{etime}
\end{deluxetable}

\clearpage

\end{landscape}

\clearpage

\clearpage

\clearpage

\end{document}